\documentclass[12pt]{article}
\usepackage{amsfonts}
\usepackage{amsmath}
\usepackage{amssymb}
\usepackage{graphicx}
\usepackage{color}
\usepackage[all, knot]{xy}
\usepackage{tikz}
\usepackage{array}
\usepackage{hyperref}
\usepackage{ulem}

\usepackage{comment}

\usepackage[utf8]{inputenc}
\usepackage{epstopdf}
\usepackage{subcaption}
\usepackage{amsthm}
\usepackage{enumitem}
\usepackage{mathrsfs}
\usepackage{mathtools}

\usepackage[margin=3cm]{geometry}

\def \be {\begin{equation}}
\def \ee {\end{equation}}
\def \bea {\begin{eqnarray}}
\def \eea {\end{eqnarray}}
\def \nn {\nonumber}

\def \rr {\raise.35ex\hbox{\small $\prime$}\kern-.17em{\mbox{\large $\imath$}}}

\def \dels {\partial\kern-.6em /\kern.1em}
\def \As {{A\kern-.5em / \kern.5em}}
\def \Ds {D\kern-.7em / \kern.5em}

\def \ks {k\kern-.5em /}
\def \ls {l\kern-.5em /}

\newcommand{\ci}[1]{}
\newcommand{\ke}{\rangle}

\newcommand{\ba}{\begin{eqnarray}}
\newcommand{\ea}{\end{eqnarray}}
\newcommand{\bal}{\begin{align}}
\newcommand{\eal}{\end{align}}
\newcommand{\bay}[1]{\left(\begin{array}{#1}}
\newcommand{\eay}{\end{array}\right)}

\newcommand{\st}[1]{|#1\ke}

\newcommand{\ket}[1]{|{#1}\rangle}

\def\xD{{\Delta}}

\def\xG{{\Gamma}}

\newcommand{\hide}[1]{}

\newlist{axioms}{enumerate}{2}
\setlist[axioms,1]{label=\textbf{A\arabic{axiomsi}.}, ref=A\arabic{axiomsi}}
\setlist[axioms,2]{label=\textbf{A\arabic{axiomsi}\rlap{\myEnumCounter{axiomsii}}.},%
                   ref=A\arabic{axiomsi}\myEnumCounter{axiomsii},%
                   align=parleft,%
                   leftmargin=0em,%
                   itemsep=1.4ex,%
                   before={\stepcounter{axiomsi}}}

  \usetikzlibrary{decorations.markings}

\begin{document}
\begin{titlepage}
%\begin{flushright}
%NORDITA 2019-102 %\\
%October,  2019
%\end{flushright}

\begin{center}

\textbf{\LARGE
CFT Dual for Timelike Geodesic in Lorentzian dS
\vskip.3cm
}
\vskip .5in
{\large
Xing Huang$^{a, b}$ \footnote{e-mail address: xingavatar@gmail.com} and
Chen-Te Ma$^{c}$ \footnote{e-mail address: yefgst@gmail.com}, 
\\
\vskip 1mm
}
{\sl
$^a$
Institute of Modern Physics, Northwest University, Xi'an 710069, China.
\\
$^b$
NSFC-SPTP Peng Huanwu Center for Fundamental Theory, Xi'an 710127, China. 
\\
$^c$
Department of Physics, Great Bay University, Dongguan, Guangdong 52300, China. 
}
\\
\vskip 1mm
\vspace{40pt}
\end{center}

%\newpage
\begin{abstract}
\noindent
We construct the Euclidean CFT$_{d}$ dual of a generic massive scalar in Lorentzian dS$_{d+1}$ via analytic continuation.
The resulting $PT$ defect defines a $PT$-invariant state that reproduces the Bunch-Davies Wightman function. 
However, the entanglement entropy captures only the real part of the central charge.
This motivates a single-geodesic dual based on the timelike geodesic-integrated Wightman function, which yields the correlators between a bulk operator and a linear combination of an OPE block and its Casimir partner.
We also derive the associated conformal defect and anomaly from an integral identity of the dS/CFT symmetry group.
\end{abstract}
\end{titlepage}
\setcounter{tocdepth}{3}
{\hypersetup{linkcolor=black}\tableofcontents}
\newpage
\section{Introduction}
\label{sec:1}
\noindent
The ($d+1$)-dimensional de Sitter (dS$_{d+1}$) spacetime is the maximally symmetric vacuum solution of Einstein's equations with a positive cosmological constant $\Lambda$.
It provides the simplest geometric model of an accelerating universe.
The cosmological constant is related to the curvature radius $L$ by
\bea
\Lambda=\frac{d(d-1)}{2L^2}.
\eea
The dS$_{d+1}$ spacetime can be realized as the hyperboloid
\bea
-X_0^2+X_{-1}^2+\sum_{j=1}^d X_j^2=L^2.
\eea
embedded in a flat spacetime with metric
\bea
ds^2=-dX_0^2+dX_{-1}^2+\sum_{j=1}^d dX_j^2.
\eea
Another maximally symmetric Lorentzian manifold is anti-de Sitter (AdS$_{d+1}$) spacetime, which is the vacuum solution of Einstein's equations with a negative cosmological constant.
It can be represented as
\bea
-X_{-1}^2-X_{0}^2+\sum_{j=1}^d X_j^2=-L^2
\eea
embedded in a flat spacetime with metric
\bea
ds^2=-dX_{-1}^2-dX_0^2+\sum_{j=1}^d dX_j^2.
\eea
Owing to the close similarity of their embedding descriptions, many results in AdS and dS spacetimes can be related through {\it analytical continuation}.
For example, the {\it global} AdS$_{d+1}$ metric
\bea
ds^2=L^2\big(-\cosh^2(\rho)d\tau^2+d\rho^2+\sinh^2(\rho)d\Omega_{d-1}^2\big)
\label{adsm}
\eea
is analytically continued by
\bea
\tau\rightarrow it; \ \rho\rightarrow i\theta; \ L^2\rightarrow-L^2
\eea
yielding
\bea
ds^2=L^2\big(-\cos^2(\theta)dt^2+d\theta^2+\sin^2(\theta)d\Omega_{d-1}^2\big),
\label{dsm}
\eea
which is the {\it static}-coordinate representation of de Sitter spacetime.
\\

\noindent
General relativity describes gravity as the curvature of spacetime.
In contrast, quantum field theory (QFT) is conventionally formulated on a fixed spacetime background.
Although the other fundamental interactions admit a successful quantum-field-theoretic description, gravity presents a unique challenge.
If one attempts to quantize Einstein gravity perturbatively in the same manner as other gauge theories, the resulting theory is non-renormalizable.
It requires an infinite number of independent counterterms.
Consequently, Einstein gravity cannot be regarded as a conventional ultraviolet (UV) complete QFT.
One possible route toward a consistent theory of quantum gravity is provided by the holographic principle, which asserts that the physical degrees of freedom contained within a spacetime region can be equivalently encoded on a lower-dimensional boundary theory \cite{tHooft:1993dmi,Susskind:1994vu}.
The most concrete realization of this idea is the AdS/CFT correspondence, where a gravitational theory in asymptotically anti-de Sitter (AdS) spacetime is dual to a conformal field theory (CFT) living on its boundary \cite{Maldacena:1997re}.
\\

\noindent
Historically, an important clue came from the asymptotic symmetry analysis of AdS$_3$ Einstein gravity.
Brown and Henneaux showed that the asymptotic symmetry algebra consists of two copies of the Virasoro algebra \cite{Brown:1986nw}, strongly suggesting that the quantum description of AdS$_3$ gravity \cite{Witten:1988hc,Witten:2007kt} is governed by a two-dimensional conformal field theory (CFT$_2$).
Since a CFT is defined at a renormalization-group fixed point, it provides a UV-complete framework for computing gravitational observables non-perturbatively.
In this sense, the dual CFT furnishes a quantum completion of the gravitational theory.
Furthermore, pure three-dimensional Einstein gravity is a topological theory and possesses no local propagating gravitational degrees of freedom.
This special property eliminates many of the difficulties associated with perturbative non-renormalizability and makes AdS$_3$/CFT$_2$ an especially tractable setting for exploring quantum gravity \cite{Witten:1988hc,Witten:2007kt}.
More broadly, holographic duality offers a framework for addressing fundamental questions of high-energy physics, including the microscopic origin of black-hole entropy, the information paradox, and the quantum nature of the early universe.
\\

\noindent
By taking the limit $\rho\to\infty$ in Eq.~\eqref{adsm}, one approaches the conformal boundary of AdS$_{d+1}$, whose geometry is $\mathbb{R}\times S^{d-1}$.
The {\it isometry group} of AdS$_{d+1}$ is SO($d$, 2), which coincides with the {\it conformal group} acting on the boundary.
This remarkable symmetry-matching provides one of the earliest pieces of evidence for the AdS/CFT correspondence: the AdS partition function is equivalent to the CFT generating function.
Functional differentiation of the CFT generating function with respect to the boundary source ($\phi_0$) then generates correlation functions of the dual CFT operators \cite{Gubser:1998bc,Witten:1998qj}.
This relation allows strongly coupled CFT observables to be computed using semiclassical gravitational calculations in the bulk.
It has led to numerous successful tests of the correspondence.
\\

\noindent
From the perspective of quantum gravity, the correspondence is particularly significant because perturbative Einstein gravity is not ultraviolet complete.
The AdS/CFT correspondence suggests that a consistent quantum theory of gravity can be formulated as a dual conformal field theory \cite{Balasubramanian:1998sn}, in which the microscopic degrees of freedom are manifest.
In this picture, locality in the bulk is not fundamental but emerges only in an appropriate limit.
Specifically, a weakly coupled semiclassical gravitational description arises when the CFT possesses a large central charge together with a sufficiently sparse spectrum of low-dimension operators.
Understanding how local bulk fields, spacetime geometry, and gravitational interactions emerge from the underlying CFT degrees of freedom remains a central goal of holography.
\\

\noindent
One necessary tool is the Hamilton-Kabat-Lifschytz-Lowe (HKLL) bulk reconstruction procedure \cite{Hamilton:2005ju,Hamilton:2006az} or the operator-product-expansion (OPE) block description through the Radon transformation \cite{Czech:2015qta,Czech:2016xec,deBoer:2016pqk,Huang:2019wzc,Huang:2020cye}.
The OPE block has the form \cite{Czech:2015qta,Czech:2016xec,deBoer:2016pqk,Huang:2019wzc,Huang:2020cye}
\bea
{\cal O}_1(x_1){\cal O}_2(x_2)=|x_{12}|^{-\Delta_1-\Delta_2}\sum_kC_{12k}B_k(x_1, x_2),
\eea
where $B_k(x_1, x_2)$ collects the contribution of the conformal family generated by the primary operator ${\cal O}_k$ of conformal dimension $\Delta_k$.
A remarkable feature of AdS/CFT is that the OPE block admits a geometric dual description \cite{Czech:2015qta,deBoer:2016pqk,Huang:2019wzc,Huang:2020cye}.
Specifically, the contribution of a conformal family can be represented by a bulk scalar field $\phi$ integrated along the geodesic connecting the boundary points $x_1$ and $x_2$ \cite{Czech:2015qta,deBoer:2016pqk,Huang:2019wzc,Huang:2020cye},
\bea
B_k(x_1, x_2) \propto \int_{\gamma_{12}}d\lambda\ \phi_k\big(Y(\lambda)\big),
\eea
where $\gamma_{12}$ denotes the bulk geodesic anchored at the boundary points $x_1$ and $x_2$.
In this correspondence, the two endpoints of the geodesic become the two insertion points appearing in the OPE block \cite{Czech:2015qta,deBoer:2016pqk,Huang:2019wzc,Huang:2020cye}.
The combination of the OPE block and the bulk scalar $B_k(x_1, x_2)\phi_k(X)$ is the bulk geodesic-integrated bulk-to-bulk propagator \cite{Hijano:2015zsa,Kabat:2016zzr}
\bea
I_{\gamma}(X)=\langle B_k(x_1, x_2)\phi_k(X)\rangle=\int_{\gamma_{12}}d\lambda \ G_{\Delta}\big(X, Y(\lambda)\big),
\eea
where
\bea
\big\langle \phi\big(Y(\lambda)\phi(X)\big\rangle=G_{\Delta}\big(X, Y(\lambda)\big).
\eea
The AdS geodesic $\gamma_{12}$ is parametrized in the $\lambda$, written in embedding space \cite{Hijano:2015zsa}
\bea
Y(\lambda)=L\frac{e^{-\lambda}P_1+e^{\lambda}P_2}{\sqrt{-2P_1\cdot P_2}},
\eea
where $P_{1, 2}$ are boundary vectors satisfy \cite{Hijano:2015zsa}
\bea
P_1^2=P_2^2=0,
\eea
and the bulk point $X$ obeys \cite{Hijano:2015zsa}
\bea
X^2=-L^2.
\eea
Because the bulk geodesic-integrated bulk-to-bulk propagator satisfies the Casimir eigenfunction with eigenvalues $\Delta(\Delta-d)$ or equivalently satisfy the bulk equation of motion \cite{Hijano:2015zsa}
\bea
(\nabla^2_{\mathrm{AdS}}-m^2)I_{\gamma}(X)=\delta_{\gamma_{12}}(X),
\eea
where the square of the bulk scalar mass is relevant to the conformal dimensions
\bea
m^2L^2=\Delta(d-\Delta).
\eea
\\

\noindent
Understanding quantum gravity requires going beyond the semiclassical regime and incorporating finite-$G_N$ effects.
In the holographic framework, such corrections correspond to finite-central-charge effects in the dual CFT.
While the AdS/CFT correspondence has passed numerous tests in the large-$c$ limit, explicit computations at finite $c$ remain considerably more challenging.
Consequently, direct evidence for the correspondence in the genuinely quantum-gravitational regime is relatively scarce.
\\

\noindent
A particularly important example is pure Einstein gravity in AdS$_3$, where significant progress has been made toward establishing an exact bulk-boundary correspondence \cite{Cotler:2018zff}.
In this setting, one can determine the relation between the three-dimensional bare gravitational coupling $G_3$ and the boundary central charge exactly at one-loop order \cite{Cotler:2018zff,Giombi:2008vd,Huang:2019nfm,Huang:2020tjl},
\bea
c=\frac{3L}{2G_3}+13.
\eea
The additive shift by 13 originates from quantum effects.
It provides a non-trivial example in which bulk and boundary parameters can be matched beyond the classical Brown--Henneaux relation.
This result demonstrates that the large-$c$ expansion of the dual CFT is naturally interpreted as a resummation of perturbative expansions in the bulk gravitational coupling \cite{Huang:2019nfm,Huang:2020tjl}.
Moreover, comparisons between bulk and boundary observables indicate that perturbative calculations organized solely in powers of the bare coupling $G_3$ should fail to reproduce the expected CFT results.
Instead, one must first incorporate the quantum shift of the central charge and reorganize the perturbative expansion in terms of the physical parameter $c$ \cite{Huang:2019nfm,Huang:2020tjl}.
This observation highlights the importance of resummation effects in establishing precise tests of the AdS/CFT correspondence beyond the semiclassical limit.
Despite these advances, genuinely non-perturbative techniques remain limited on both sides of the duality.
It is therefore still an open question to what extent the CFT provides a complete microscopic description of quantum gravity.
Nevertheless, holography offers a unique framework in which spacetime locality, gravitational dynamics, and semiclassical geometry emerge from a fundamentally non-gravitational quantum system.
\\

\noindent
Since the static slicing of dS spacetime can be obtained from global AdS through an analytical continuation, it is natural to explore a holographic correspondence between de Sitter spacetime and a conformal field theory by analytically continuing the AdS/CFT framework \cite{Strominger:2001pn}.
In the dS/CFT correspondence, the dual Euclidean CFT is conjectured to be associated with the asymptotic {\it future} and {\it past} boundaries corresponding to $t=\pm\infty$ in Eq. \eqref{dsm}  \cite{Anninos:2011ui}.
A simple indication of the difference between AdS/CFT and dS/CFT arises from the analytical continuation of the AdS curvature radius.
Applying $L\rightarrow iL$ to the holographic relation between the central charge and the gravitational coupling yields \cite{Strominger:2001pn}
\bea
c\sim \frac{L^{d-1}}{G_N}\ \longrightarrow i^{d-1}\frac{L^{d-1}}{G_N}.
\eea
For $d=2$, this suggests that the dual CFT possesses a complex central charge.
Such a theory cannot be interpreted as a conventional unitary CFT and therefore {\it differs} fundamentally from the boundary theories in standard AdS/CFT.
This distinction becomes more explicit when one examines the realization of the conformal generators.
In the AdS$_3$ bulk-coordinate representation, the global conformal generators satisfy
\bea
\lbrack L_m, L_n\rbrack=(m-n)L_{m+n}; \ \lbrack \tilde{L}_m, \tilde{L}_n\rbrack=(m-n)\tilde{L}_{m+n}
\label{gva}
\eea
with $m, n=-1, 0, 1$, together with the conventional Hermitian conjugation relations
\bea
(L_n)^{\dagger}=L_{-n}; \ (\tilde{L}_n)^{\dagger}=\tilde{L}_{-n}.
\eea
After analytically continuing from AdS$_3$ to dS$_3$, the generators continue to satisfy the same algebra, but the adjoint operation is modified to \cite{Doi:2024nty}
\bea
&&
(L_0)^{\dagger}=-\tilde{L}_{0}, \ (L_{\pm 1})^{\dagger}=\tilde{L}_{\pm 1};
\nn\\
&&
(\tilde{L}_0)^{\dagger}=-L_{0}; \ (\tilde{L}_{\pm 1})^{\dagger}=L_{\pm 1}.
\eea
Thus, while the algebraic structure remains sl(2, $\mathbb{R}$)$\oplus$ sl(2, $\mathbb{R}$), the associated reality condition is fundamentally altered.
The resulting theory admits a {\it non-standard} inner product.
It provides evidence that a {\it non-unitary} conformal field theory governs the dS/CFT correspondence.
This {\it unusual} adjoint structure plays a central role in the analytically continued dS/CFT framework \cite{Doi:2024nty}.
\\

\noindent
Since the modified adjoint operation is determined solely by the realization of the de Sitter isometry group as the conformal group on the boundary, the construction can be generalized from dS$_3$/CFT$_2$ to arbitrary dimensions \cite{Huang:2025gmq}.
For a CFT$_d$ dual to dS$_{d+1}$, the conformal generators satisfy the non-standard adjoint relations \cite{Huang:2025gmq}
\bea
D^{\dagger}=-D; \
K_{\mu}^{\dagger}=K_{\mu}; \
P_{\mu}^{\dagger}=P_{\mu},
\eea
where $D$, $K_{\mu}$, and $P_{\mu}$ generate the dilatations, special conformal transformations, and translations, respectively.
These generators continue to obey the standard conformal algebra.
However, the associated reality condition differs from that of a unitary Euclidean CFT.
\\

\noindent
When the conformal generators act on the primary states $\ket{\Delta_\pm}$, which are the lowest-weight states, the adjoint operation does not have the conventional form, and it depends on the value of the mass.
We consider a real scalar field with a given mass $m$, and then use the analytical continuation from the AdS$_{d+1}$ case to obtain two conformal dimensions for the dS$_{d+1}$ case
\bea
\Delta_{\pm}=\frac{d}{2}\pm\sqrt{\frac{d^2}{4}-m^2L^2}\equiv\frac{d}{2}\pm\mu.
\eea
When considering the heavy mass
\bea
m^2L^2>\frac{d^2}{4},
\label{hm}
\eea
the results for the adjoint operation when the conformal generators act on the primary states are \cite{Huang:2025gmq}
\bea
&&
\langle\hat{\Delta}_{\pm}|D^{\dagger}=\langle\Delta_{\pm}|\hat{\Delta}_{\mp}; \
D^{\dagger}|\hat{\Delta}_{\pm}\rangle=\Delta_{\mp}|\hat{\Delta}_{\pm}\rangle; \
0=\langle\hat{\Delta}_{\pm}|K_{\mu}^{\dagger}=P^{\dagger}_{\mu}|\hat{\Delta}_{\pm}\rangle;
\nn\\
&&
\langle\hat{\Delta}_{\pm}|D=-\langle\hat{\Delta}_{\pm}|\Delta_{\mp}; \
D|\hat{\Delta}_{\pm}\rangle=-\Delta_{\mp}|\hat{\Delta}_{\pm}\rangle; \
0=\langle\hat{\Delta}_{\pm}|K_{\mu}=P_{\mu}|\hat{\Delta}_{\pm}\rangle,
\eea
where the normalziation is \cite{Huang:2025gmq} :
\bea
\langle\Delta_j|\Delta_k\rangle=\langle\hat{\Delta}_j|\hat{\Delta}_k\rangle=\delta_{jk}; \
\langle\Delta_j|\hat{\Delta}_k\rangle=\langle\hat{\Delta}_j|\Delta_k\rangle=0,
\eea
and the adjoint bra state as \cite{Huang:2025gmq}:
\bea
(|\Delta_{\pm}\rangle)^{\dagger}\equiv\langle\hat{\Delta}_{\pm}|; \
(\langle\Delta_{\pm}|)^{\dagger}=|\hat{\Delta}_{\pm}\rangle.
\eea
Thus, unlike ordinary unitary conformal field theories, the de Sitter realization requires a {\it doubled} state space together with a non-standard adjoint operation.
The resulting representation {\it modifies} the notion of Hermiticity and inner product, providing the foundation for the $PT$-symmetric structure \cite{Bender:1998ke,Bender:2002vv,Mostafazadeh:2001jk,Mostafazadeh:2001nr,Mostafazadeh:2002id,Mostafazadeh:2002maq} in the dS/CFT correspondence, where $P$ and $T$ are the parity and the time-reversal transformations of the bulk coordinates \cite{Doi:2024nty,Huang:2025gmq}.
\\

\noindent
The analytical continuation from global AdS does not yield the conventional single-static patch of de Sitter spacetime.
Instead, it yields a doubled geometry in which the angular coordinate extends over the range
\bea
0<\theta<\pi
\eea
rather than the usual static-patch interval ($0<\theta<\pi/2$) \cite{Doi:2024nty,Huang:2025gmq}.
Consequently, the analytically continued spacetime contains two static regions related by an {\it antipodal} transformation.
\\

\noindent
The antipodal map is generated by $\exp(i\theta J)$ \cite{Doi:2024nty,Huang:2025gmq}, where
\bea
J\equiv\frac{1}{2}(K_0-P_0),
\eea
and acts geometrically as \cite{Doi:2024nty,Huang:2025gmq}
\bea
\theta \rightarrow \theta+\pi.
\eea
On a sphere, antipodal points are connected through a diameter passing through the center of the sphere \cite{Doi:2024nty,Huang:2025gmq}.
The doubled static-patch geometry, therefore, naturally incorporates an additional $\mathbb{Z}_2$ symmetry associated with the antipodal identification \cite{Doi:2024nty,Huang:2025gmq}.
\\

\noindent
The Green's function (Wightman function) for a real scalar field in a {\it Bunch-Davies vacuum} \cite{Bousso:2001mw}
\bea
G_E(x, y)=\frac{\Gamma(\Delta_+)\Gamma(\Delta_-)}{(4\pi)^{\frac{d+1}{2}}\Gamma\big(\frac{d+1}{2}\big)}
{}_2F_1\bigg\lbrack\Delta_+, \Delta_-, \frac{d+1}{2}; \cos^2\bigg(\frac{D_{\mathrm{dS}}(x, y)}{2}\bigg)\bigg\rbrack,
\eea
where $D_{\mathrm{dS}}$ is the geodesic for the dS$_{d+1}$ background, is compatible with the antipodal symmetry and therefore admits a natural $PT$-symmetry interpretation \cite{Doi:2024nty,Huang:2025gmq}.
This observation allows one to formulate a $PT$ inner product \cite{Bender:1998ke,Bender:2002vv,Mostafazadeh:2001jk,Mostafazadeh:2001nr,Mostafazadeh:2002id,Mostafazadeh:2002maq} for bulk local states (bulk scalar field acting on a vacuum state) \cite{Doi:2024nty,Huang:2025gmq,Ishibashi:1988kg,Nakayama:2015mva}
\bea
\label{bklocal}
|\Psi_{\Delta}\rangle= \sum_{n=0}^{\infty} (-1)^n C_n (P^2)^n | \Delta \rangle,
\eea
where
\bea
P^2 = \sum_{a=0}^{d-1} P_a P_a; \
C_n = \prod_{k=1}^n \frac{1}{4k \Delta + 4 k^2 -2kd},
\eea
and $|\Delta\rangle$ denotes a conformal primary state.
This state satisfies \cite{Ishibashi:1988kg,Nakayama:2015mva}
\bea
(P_\mu + K_\mu)|\Psi_\Delta\rangle =  0.
\label{blsv}
\eea
The Bunch-Davies Wightman function can then be reproduced from the overlap of such bulk local states, evaluated using the $PT$ inner product, in the {\it heavy}-mass regime \cite{Doi:2024nty,Huang:2025gmq}.
This equivalence suggests that the analytically continued dS/CFT correspondence naturally admits a $PT$-symmetric Hilbert-space structure.
Although an HKLL reconstruction of bulk fields in de Sitter space can be formulated \cite{Xiao:2014uea}, a complete analog of the AdS Radon-transform/OPE-block correspondence has not yet been established \cite{Nath:2024aqh}.
Nevertheless, the doubled static-patch geometry shares several structural features with AdS and provides a natural setting in which to study conformal defects \cite{Billo:2016cpy} and their associated integral identities \cite{Huang:2025gmq,Belton:2025ief,Drukker:2025dfm}.
\\

\noindent
The existence of a $PT$ inner product introduces an additional $\mathbb{Z}_2$ symmetry that proves useful in the analysis of conformal defects and anomalies \cite{Huang:2025gmq,Belton:2025ief,Drukker:2025dfm}.
However, for CFT$_2$ theories with {\it complex} central charge, the vacuum state spontaneously breaks the $PT$ symmetry  \cite{Huang:2025gmq}.
Once a Hilbert-space structure is established, one can further apply {\it replica} techniques to investigate entanglement measures and extract information about the central charge \cite{Holzhey:1994we}.
Comparing these quantities with their AdS counterparts provides a new avenue for understanding the geometric interpretation of quantum information in de Sitter holography \cite{Ryu:2006bv,Ryu:2006ef,Lewkowycz:2013nqa}.
Thus, the construction of a $PT$-symmetric Hilbert space opens the door to studying quantum-information aspects \cite{Nakata:2020luh,Doi:2022iyj} of the dS/CFT correspondence.

\subsection{Summary of Results}
\noindent
In this paper, we mainly investigate the Euclidean CFT dual of a scalar field integrated along timelike geodesics in the Lorentzian de Sitter space.
The timelike geodesic follows from an antipodal map that relates the endpoints of each boundary.
We find that the geodesic operator from integration contains CFT information, namely the OPE blocks of both conformal dimensions $\xD_\pm$.
To get quantities like the Wightman function or the entanglement entropy, a certain linear combination required by $PT$ symmetry is needed.
We expect that these results should point toward a new type of model on dS/CFT correspondence involving CFTs on both boundaries with $PT$ symmetry imposing a constraint.
Our main results are summarized as follows:
\begin{itemize}
\item Previous studies of the analytically continued dS/CFT correspondence have primarily focused on the principal-series regime \big(Eq. \eqref{hm}\big), where the scalar field is sufficiently massive.
The conformal dimensions form a complex conjugate pair.
In contrast, we extend the construction to the complementary-series regime, where
\bea
(\Delta_{\pm})^*=\Delta_{\pm},
\eea
and the conformal dimensions are real.
Since the $PT$ transformation no longer exchanges the two conformal weights, the corresponding $PT$-invariant states differ significantly from those in the heavy-mass regime. Understanding this structure is essential for obtaining a complete picture of the analytically continued dS/CFT correspondence. We construct the $PT$-invariant states in the complementary-series regime and show that they reproduce the Bunch--Davies two-point Wightman function.
\item We compute the entanglement entropy of a CFT$_2$ with complex central charge and identify its bulk dual in terms of the $PT$-dual geodesic.
We find that the resulting entropy depends only on the real part of $c$.
\item We show that the corresponding geodesic-integrated Wightman function is dual to the correlators between a bulk operator and a linear combination of an OPE block of conformal dimension $\xD_+$ and its partner with conformal dimension $\xD_-$ beginning from the static patch and then undergoing an analytical continuation to obtain the dual for the global patch as well.
This construction provides a bulk observable sensitive to the imaginary part of $c$.
\item We further discuss conformal defects, the associated anomaly derived from an integral identity, and their possible interpretation in the bulk.
\end{itemize}

\noindent
The remainder of this paper is organized as follows.
In Sec.~\ref{sec:2}, we analyze the action of the conformal generators on the primary states $|\Delta_\pm\rangle$ in the complementary-series regime.
In Sec.~\ref{sec:3}, we construct the bulk local states relevant for the analytic continuation of the Wightman function.
In Sec.~\ref{sec:4}, we introduce the global $PT$-defect operator and show how the $PT$-twisted inner product reproduces the de Sitter two-point function.
In Sec.~\ref{sec:5}, we compute the entanglement entropy of CFT$_2$ with complex central charge and identify its corresponding bulk dual.
In Sec.~\ref{sec:6}, we establish the duality between the geodesic-integrated Wightman function and a linear combination of an OPE block and its partner.
In Sec.~\ref{sec:7}, we discuss conformal defects, the associated anomaly obtained from an integral identity, and their possible bulk interpretation.
Finally, Sec.~\ref{sec:8} contains our conclusions and future directions.
We derive some useful identities that show that the geodesic-integrated Wightman function satisfies the bulk equation of motion in Appendix~\ref{sec:A}.

%The main results of our work are presented in Fig. \ref{summary.png}.
%\begin{figure}[tbp]
%\centering
%\includegraphics[scale = 1]{fig-Z23.pdf}
%\caption{Graphical notation for the tripartite wavefunction $\Psi$, its conjugate $\Psi^*$, the reduced density matrix $\rho_{BC}$, and the construction of $\mathcal{Z}^{(\mathtt{q})}_{2}$ in the special case $\mathtt{q} = 3$. }
%\label{fig:Z23}
%\end{figure}

\section{Conformal Generators}
\label{sec:2}
\noindent
For the principal-series regime, the conformal dimensions are the complex numbers with
\bea
(\Delta_+)^*=\Delta_-.
\eea
Therefore, the complex conjugation exchanges the conformal dimensions. However, it does not in the complementary-series regime because the conformal dimensions are real-valued.
This generates the difference to the complementary-series regime when the conformal generators act on the primary states.
The result for the complementary series regime is
\bea
&&
\langle\hat{\Delta}_{\pm}|D^{\dagger}=\langle\Delta_{\pm}|\hat{\Delta}_{\pm}; \
D^{\dagger}|\hat{\Delta}_{\pm}\rangle=\Delta_{\pm}|\hat{\Delta}_{\pm}\rangle; \
0=\langle\hat{\Delta}_{\pm}|K_{\mu}^{\dagger}=P^{\dagger}_{\mu}|\hat{\Delta}_{\pm}\rangle;
\nn\\
&&
\langle\hat{\Delta}_{\pm}|D=-\langle\hat{\Delta}_{\pm}|\Delta_{\pm}; \
D|\hat{\Delta}_{\pm}\rangle=-\Delta_{\pm}|\hat{\Delta}_{\pm}\rangle; \
0=\langle\hat{\Delta}_{\pm}|K_{\mu}=P_{\mu}|\hat{\Delta}_{\pm}\rangle.
\eea

\section{Bulk Local State}
\label{sec:3}
\noindent
We can apply $\exp(Dt){\cal R}\exp(i\theta J)$ to a bulk local state, and then it takes any bulk point to anywhere on a sphere, $S^{d-1}$.
Because we have the following identity
\bea
e^{i\pi J}De^{-i\pi J}=-D,
\eea
we obtain
\bea
\langle\hat{\Delta}_{\pm}|e^{i\pi J}D=\Delta_{\pm}\langle\hat{\Delta}_{\pm}|e^{i\pi J},
\eea
whcih shows that $\langle\hat{\Delta}_{\pm}|e^{i\pi J}$ is proportional to a conjugate transpose of the primary state $\langle\Delta_{\pm}|$,
\bea
\langle\hat{\Delta}_{\pm}|e^{i\pi J}=\nu_{\pm}\langle\Delta_{\pm}|.
\eea
We take the conjugate transpose to show
\bea
e^{-i\pi J}|\Delta_{\pm}\rangle=\nu_{\pm}^*|\hat{\Delta}_{\pm}\rangle.
\eea
Hence, we get
\bea
\nu_{\pm}\langle\Delta_{\pm}|e^{-2\pi J}|\Delta_{\pm}\rangle=\nu_{\pm}^*.
\eea
The $\exp(i\pi J) \st{\Psi_{\xD}}$ remains a bulk local state, and then we use $\exp(D t) {\cal R}$ to take a dS bulk point to anywhere. These operations provide the following equality
\bea
|\Psi_{\Delta_{\pm}}(t, \theta+\pi, \Omega)\rangle=\lambda_{\pm}|\Psi_{\Delta_{\pm}}(t, \theta, \Omega)\rangle.
\eea
The bulk local state \eqref{bklocal} can be carried out and gives
\bea
\label{bulklocald}\st{\Psi_\Delta} = \xG\bigg(\Delta-\frac{d}{2}+1\bigg) \left(\frac{\sqrt{P^2}}{2}\right)^{\frac{d}{2}-\Delta} J_{\Delta-\frac{d}{2}} (\sqrt{P^2})|\xD \rangle.
\eea
\\

\noindent
The bulk-to-bulk propagator for a real scalar in the AdS case is
\bea
G(x, y)=\frac{\Gamma(\Delta)}{2\pi^{\frac{d}{2}}\Gamma\big(\Delta-\frac{d}{2}+1\big)}
e^{-\Delta D_{\mathrm{AdS}}(x, y)} {}_2F{}_1\bigg(\Delta, \frac{d}{2}, \Delta+1-\frac{d}{2}; e^{-2D_{\mathrm{AdS}}(x, y)}\bigg),
\eea
where the Gamma function is
\bea
\Gamma(z)\equiv\int_0^{\infty}dt\ t^{z-1}e^{-t}, \ \mathrm{Re}(z)>0,
\eea
and the hypergeometric function for $|z|<1$ can be defined by the infinite sum of a series
\bea
{}_2F_1(a, b, c; z)\equiv\sum_{n=0}^{\infty}\frac{(a)_n(b)_n}{(c)_n}\frac{z^n}{n!}, \ |z|<1.
\eea
The Pochhammer symbol is defined as
\bea
(a)_n\equiv\left\{\begin{array}{ll}
1, & \mbox{$n=0$}. \\
a(a+1)\cdots (a+n-1), & \mbox{$n>0$}.
\end{array} \right.
\eea
The $D_{\mathrm{AdS}}(x, y)$ is the geodesic for the AdS$_{d+1}$ background between two bulk points, $x$ and $y$.
For $|z|\ge 1$, we apply the analytical continuation to the hypergeometric function or use other representations to compute it.
However, the series cannot be applied to the hypergeometric function anymore for $|z|\ge 1$.
We then use the analytical continuation to obtain the following Green's function
\bea
G_{\Delta_{\pm}}(x, y)=\frac{\Gamma(\Delta_{\pm})}{2\pi^{\frac{d}{2}}\Gamma\big(\Delta_{\pm}-\frac{d}{2}+1\big)}
e^{-i\Delta_{\pm} D_{\mathrm{dS}}(x, y)} {}_2F{}_1\bigg(\Delta_{\pm}, \frac{d}{2}, \Delta_{\pm}+1-\frac{d}{2}; e^{-2iD_{\mathrm{dS}}(x, y)}\bigg),
\label{gf1}
\nn\\
\eea
where $D_{\mathrm{dS}}(x, y)$ is the geodesic line for the dS$_{d+1}$ background connecting two bulk points $x$ and $y$, and it connectes to $D_{\mathrm{AdS}}(x, y)$ as
\bea
D_{\mathrm{AdS}}(x, y)=iD_{\mathrm{dS}}(x, y).
\eea
According to Eq. \eqref{gf1}, we can read
\bea
\frac{1}{\lambda_{\pm}}=(-1)^{\frac{d}{2}}e^{\mp i\pi\mu}
\eea
from:
\bea
\langle\Psi_{\Delta_{\pm}}(0, \theta+\pi, 0)|\Psi_{\Delta_{\pm}}(0, 0, 0)\rangle
&=&\langle\Psi_{\Delta_{\pm}}(0, \theta, 0)|  e^{-i\pi J} |\Psi_{\Delta_{\pm}}(0, 0, 0)\rangle
\nn\\
&=&(-1)^{\frac{d}{2}}e^{\mp i\pi\mu}\langle\Psi_{\Delta_{\pm}}(0, \theta, 0)|\Psi_{\Delta_{\pm}}(0, 0, 0)\rangle.
\eea
Hence, we can show
\bea
\langle\Delta_{\pm}|e^{-2\pi J}|\Delta_{\pm}\rangle=(-1)^de^{\mp 2i\pi\mu},
\eea
and
\bea
\nu_{\pm}=(-1)^{\frac{d}{2}+1}e^{\pm i\pi\mu}.
\eea
\\

\noindent
We can take the conjugate transpose of the bulk local state to get:
\bea
\langle\hat{\Psi}_{\Delta_{\pm}}|&=&
\bigg\langle\hat{\Delta}_{\pm}\bigg|\sum_{n=0}^{\infty}(-1)^nC_n(\Delta_{\pm})(P^2)^ne^{-i\theta J}{\cal R}^{-1}e^{-Dt}
\nn\\
&=&
\bigg\langle\hat{\Delta}_{\pm}\bigg|\sum_{n=0}^{\infty}(-1)^nC_n(\Delta_{\pm})(P^2)^ne^{i\pi J}e^{-i\pi J}
e^{-i\theta J}{\cal R}^{-1}e^{-Dt}
\nn\\
&=&\bigg\langle \hat{\Delta}_{\pm}\bigg|e^{i\pi J}\sum_{n=0}^{\infty}(-1)^nC_n(\Delta_{\pm})(K^2)^ne^{-i(\theta+\pi) J}{\cal R}^{-1}e^{-Dt}
\nn\\
&=&\nu_{\pm}\bigg\langle \Delta_{\pm}\bigg|\sum_{n=0}^{\infty}(-1)^nC_n(\Delta_{\pm})(K^2)^ne^{-i(\theta+\pi) J}{\cal R}^{-1}e^{-Dt}
\nn\\
&=&\nu_{\pm}\langle\Psi_{\Delta_{\pm}}(t, \theta+\pi, \Omega)|.
\eea
Therefore, we show that the inner product of the non-vanishing bulk local states is
\bea
&&
\langle\hat{\Psi}_{\Delta_{\pm}}(x)|\Psi_{\Delta_{\pm}}(y)\rangle
\nn\\
&=&\frac{\nu_{\pm}\Gamma(\Delta_{\pm})}{2\pi^{\frac{d}{2}}\Gamma\big(\Delta_{\pm}-\frac{d}{2}+1\big)}
e^{-i\Delta_{\pm} D_{\mathrm{dS}}(x_A, y)} {}_2F{}_1\bigg(\Delta_{\pm}, \frac{d}{2}, \Delta_{\pm}+1-\frac{d}{2}; e^{-2iD_{\mathrm{dS}}(x_A, y)}\bigg),
\eea
where $x_A$ is the antipodal point, with respect to the dS bulk point $x$, and the other inner product of the bulk local states vanishes
\bea
\langle\hat{\Psi}_{\Delta_{\pm}}(x)|\Psi_{\Delta_{\mp}}(y)\rangle=0.
\eea
In the next section, we will make a linear combination of the bulk local states $|\Psi_{\Delta_{\pm}}\rangle$ to obtain the Wightman function for the real scalar field.

\section{Green's Function}
\label{sec:4}
\noindent
The Wightman function for a real scalar field in a Bunch-Davies vacuum state takes the following form \cite{Bousso:2001mw}
\bea
G_E(x, y)=\frac{\Gamma(\Delta_+)\Gamma(\Delta_-)}{(4\pi)^{\frac{d+1}{2}}\Gamma\big(\frac{d+1}{2}\big)}
{}_2F_1\bigg\lbrack\Delta_+, \Delta_-, \frac{d+1}{2}; \cos^2\bigg(\frac{D_{\mathrm{dS}}(x, y)}{2}\bigg)\bigg\rbrack.
\eea
We use the following identities:
\bea
\cos^2\bigg(\frac{D_{\mathrm{dS}}(x_A, y)}{2}\bigg)
=\sin^2\bigg(\frac{D_{\mathrm{dS}}(x, y)}{2}\bigg)
=1-\cos^2\bigg(\frac{D_{\mathrm{dS}}(x, y)}{2}\bigg),
\eea
where
\bea
D_{\mathrm{dS}}(x_A, y)=\pi-D_{\mathrm{dS}}(x, y),
\eea
\bea
{}_2F_1\bigg(a, b, \frac{a+b+1}{2}; 1-z\bigg)
={}_2F_1\bigg(\frac{a}{2}, \frac{b}{2}, \frac{a+b+1}{2}; 4z(1-z)\bigg)
\eea
with
\bea
z=\cos^2\bigg(\frac{D_{\mathrm{dS}}(x, y)}{2}\bigg)
\eea
to show
\bea
G_E(x_A, y)=\frac{\Gamma(\Delta_+)\Gamma(\Delta_-)}{(4\pi)^{\frac{d+1}{2}}\Gamma\big(\frac{d+1}{2}\big)}
{}_2F_1\bigg(\frac{\Delta_+}{2}, \frac{\Delta_-}{2}, \frac{d+1}{2}; \sin^2\big(D_{\mathrm{dS}}(x, y)\big)\bigg).
\eea
We then apply another identity
\bea
{}_2F_1\bigg(a, b, 2b; z\bigg)=(1-z)^{-\frac{a}{2}}{}_2F_1\bigg(\frac{a}{2}, b-\frac{a}{2}, b+\frac{1}{2}; \frac{z^2}{4(z-1)}\bigg)
\eea
to rewrite $G_E(x_A, y)$ as
\bea
G_E(x_A, y)=\frac{\Gamma(\Delta_+)\Gamma(\Delta_-)}{(4\pi)^{\frac{d+1}{2}}\Gamma\big(\frac{d+1}{2}\big)}
e^{i\Delta_+ D_{\mathrm{dS}}}
{}_2F_1\bigg(\Delta_+, \frac{d}{2}, d; 1-e^{2iD_{\mathrm{dS}}}\bigg).
\eea
The $G_E(x_A, y)$ can also be rewritten as the symmetric form between $\Delta_+$ and $\Delta_-$,
\bea
&&
G_E(x_A, y)
\nn\\
&=&\frac{\Gamma(\Delta_-)\Gamma\big(\frac{d}{2}-\Delta_-)}{4(\pi)^{\frac{d}{2}+1}}
e^{i\Delta_-D_{\mathrm{dS}}(x, y)}
{}_2F_1\bigg(\Delta_-, \frac{d}{2}, -\frac{d}{2}+\Delta_-+1; e^{2i D_{\mathrm{dS}}(x, y)}\bigg)
\nn\\
&&+\frac{\Gamma(\Delta_+)\Gamma\big(\frac{d}{2}-\Delta_+)}{4(\pi)^{\frac{d}{2}+1}}
e^{i\Delta_+D_{\mathrm{dS}}(x, y)}
{}_2F_1\bigg(\Delta_+, \frac{d}{2}, -\frac{d}{2}+\Delta_++1; e^{2i D_{\mathrm{dS}}(x, y)}\bigg),
\nn\\
\eea
by using the following formula
\bea
&&
{}_2F_1(a, b, c; z)
\nn\\
&=&\frac{(1-z)^{-a-b+c}\Gamma(c)\Gamma(a+b-c)}{\Gamma(a)\Gamma(b)}
{}_2F_1(c-a, c-b; -a-b+c+1; 1-z)
\nn\\
&&
+\frac{\Gamma(c)\Gamma(-a-b+c)}{\Gamma(c-a)\Gamma(c-b)}
{}_2F_1(a, b, a+b-c+1; 1-z).
\eea
\\

\noindent
According to the symmetric form, we can use the linear combination of the primaries $\Delta_{\pm}$ to write the dual wavefunction
\bea
&&
|\Psi_E(x)\rangle
\nn\\
&=&\frac{1}{\sqrt{2\sinh(-i\pi\mu)}}\bigg\lbrack\bigg(\frac{1}{2}-\frac{i}{2}\bigg)|\Psi_{\Delta_+}(x)\rangle
+\bigg(\frac{1}{2}-\frac{i}{2}\bigg)|\Psi_{\Delta_+}(x_A)\rangle
\nn\\
&&
-\bigg(\frac{1}{2}+\frac{i}{2}\bigg)
|\Psi_{\Delta_-}(x)\rangle
+\bigg(\frac{1}{2}+\frac{i}{2}\bigg)|\Psi_{\Delta_-}(x_A)\rangle
\bigg\rbrack,
\nn\\
\eea
and then $G_E(x, y)$ can be obtained by replacing $x$ with $x_A$ in the $D_{\mathrm{dS}}(x, y)$ from $G_E(x_A, y)$ and can be written as an inner product:
\bea
G_E(x, y)
&=&\langle\Psi_E(x)|\Psi_E(y)\rangle
\nn\\
&=&\frac{i}{2\sinh(-i\pi\mu)}
\bigg(\langle\hat{\Psi}_{\Delta_-}(x)|\Psi_{\Delta_-}(y_A)\rangle
-\langle\hat{\Psi}_{\Delta_+}(x_A)|\Psi_{\Delta_+}(y)\rangle\bigg),
\nn\\
\eea
where
\bea
&&
\langle\Psi_E(x)|
\nn\\
&=&\frac{1}{\sqrt{2\sinh(-i\pi\mu)}}\bigg\lbrack\bigg(\frac{1}{2}-\frac{i}{2}\bigg)\langle\hat{\Psi}_{\Delta_+}(x)|
+\bigg(\frac{1}{2}-\frac{i}{2}\bigg)\langle\hat{\Psi}_{\Delta_+}(x_A)|
\nn\\
&&
+\bigg(\frac{1}{2}+\frac{i}{2}\bigg)\langle\hat{\Psi}_{\Delta_-}(x)|
-\bigg(\frac{1}{2}+\frac{i}{2}\bigg)\langle\hat{\Psi}_{\Delta_-}(x_A)
|\bigg\rbrack.
\eea
\\

\noindent
Finally, let us discuss the inner product motivated by the linear combination of the wavefunction.
The bra state can be obtained from the $PT$ transformation
\bea
\nu_{+}\langle 0|PT(PT)^{-1}\Phi_{\Delta_+}(t, \theta, \Omega)PT
=\nu_{+}\langle 0|_{PT}\Phi_{\Delta_+}(t, \theta+\pi, \Omega)
=\langle\hat{\Psi}_{\Delta_+}(x)|,
\eea
where
\bea
\langle 0|_{PT}\equiv\langle 0|PT.
\eea
The $\nu_{\pm}$ can be thought of as the linear combination of the coefficients for the states
\bea
&&
\langle\Psi_E(x)|
\nn\\
&=&\frac{1}{\sqrt{2\sinh(-i\pi\mu)}}\bigg\lbrack\bigg(\frac{1}{2}-\frac{i}{2}\bigg)\nu_+\langle\hat{\tilde{\Psi}}_{\Delta_+}(x)|
+\bigg(\frac{1}{2}-\frac{i}{2}\bigg)\nu_+\langle\hat{\tilde{\Psi}}_{\Delta_+}(x_A)|
\nn\\
&&
+\bigg(\frac{1}{2}+\frac{i}{2}\bigg)\nu_-\langle\hat{\tilde{\Psi}}_{\Delta_-}(x)|
-\bigg(\frac{1}{2}+\frac{i}{2}\bigg)\nu_-\langle\hat{\tilde{\Psi}}_{\Delta_-}(x_A)
|\bigg\rbrack,
\eea
where
\bea
\nu_{\pm}\langle\hat{\tilde{\Psi}}_{\Delta_{\pm}}|=\langle\hat{\Psi}_{\Delta_{\pm}}|.
\eea
The conventional $PT$ inner product also helps define $\langle\hat{\tilde{\Psi}}_{\Delta_{\pm}}|$,
\bea
\langle\hat{\tilde{\Psi}}_{\Delta_{\pm}}(x_A)|=\langle\Psi_{\Delta_{\pm}}(x)|PT.
\eea
It is equivalent to twisting the inner product space by the $PT$ transformation \cite{Bender:1998ke,Bender:2002vv,Mostafazadeh:2001jk,Mostafazadeh:2001nr,Mostafazadeh:2002id,Mostafazadeh:2002maq}, which equivalently maps a dS bulk point to an antipodal point.
The $PT$ operation acting on the $|\Psi_E\rangle$ is invariant due to that
\bea
&&
PT\bigg\lbrack\frac{1}{\sqrt{2\sinh(-i\pi\mu)}}\bigg(\frac{1}{2}-\frac{i}{2}\bigg)|\Psi_{\Delta_+}(x)\rangle\bigg\rbrack
\nn\\
&&
\longleftrightarrow
PT\bigg\lbrack\frac{1}{\sqrt{2\sinh(-i\pi\mu)}}\bigg(\frac{1}{2}-\frac{i}{2}\bigg)|\Psi_{\Delta_+}(x_A)\rangle\bigg\rbrack
\nn\\
&&
PT\bigg\lbrack\frac{1}{\sqrt{2\sinh(-i\pi\mu)}}\bigg(\frac{1}{2}+\frac{i}{2}\bigg)
|\Psi_{\Delta_-}(x)\rangle\bigg\rbrack
\nn\\
&&
\longleftrightarrow
PT\bigg\lbrack-\frac{1}{\sqrt{2\sinh(-i\pi\mu)}}\bigg(\frac{1}{2}+\frac{i}{2}\bigg)|\Psi_{\Delta_-}(x_A)\rangle
\bigg\rbrack.
\eea
The linear combination yields an $PT$-invariant state.
\\

\noindent
The $PT$ operator is a global defect and acts as a symmetry of the theory \cite{Huang:2025gmq}.
Therefore, if we do not have spontaneous symmetry breaking, the global defect acting on a vacuum state generates the same vacuum state, up to the coefficient \cite{Huang:2025gmq}.
However, we know that CFT$_2$ with the non-zero imaginary part of the central charge has the spontaneous $PT$-symmetry breaking in the vacuum state \cite{Huang:2025gmq}.
Indeed, this result corresponds to the antipodal symmetry.
For any dimension, the sphere has antipodal symmetry.
Hence, we expect that the spontaneous $PT$-symmetry breaking is generic for the CFT's vacuum state when the central charge has the imaginary part \cite{Huang:2025gmq}.

\section{Entanglement Entropy}
\label{sec:5}
\noindent
We consider the $PT$-invariant state
\bea
|\Psi\rangle=\frac{1}{\sqrt{N}}(|0\rangle+|\tilde{0}\rangle)
\eea
from the two $PT$-related vacuum states \cite{Huang:2025gmq}
\bea
|\tilde{0}\rangle\equiv PT|0\rangle.
\eea
The normalization factor is
\bea
N\equiv 2+2\mathrm{Re}\big(\langle 0|\tilde{0}\rangle\big).
\eea
Using this $PT$-invariant state, we compute the entanglement entropy of CFT$_2$ for a single interval.
The result shows the loss of the imaginary part of the central charge, equivalent to losing the classical gravity term.
Hence, the entanglement entropy can only affect the real part of the central charge.
We also use the $PT$-pair of geodesics obtained via analytical continuation, which yields a consistent result.

\subsection{Single Interval in CFT$_2$}
\noindent
For a single interval $A$, the reduced density matrix is:
\bea
\rho_A\equiv\mathrm{Tr}_{\bar{A}}|\Psi\rangle\langle\Psi|=\frac{1}{N}(\rho_A^{(0)}+\rho_A^{(\tilde{0})}+\sigma_A+\sigma_A^{\dagger}),
\eea
where
\bea
\rho_A^{(0)}\equiv\mathrm{Tr}_{\bar{A}}|0\rangle\langle 0|; \
\rho_A^{(\tilde{0})}\equiv\mathrm{Tr}_{\bar{A}}|\tilde{0}\rangle\langle\tilde{0}|; \
\sigma_A\equiv\mathrm{Tr}_{\bar{A}}|0\rangle\langle\tilde{0}|.
\eea
The entanglement entropy is
\bea
S_A(\Psi)\equiv -\mathrm{Tr}\big(\rho_A(\Psi)\ln\rho_A(\Psi)\big).
\eea
For a single interval
\bea
A=\lbrack u, v\rbrack
\eea
in a 2D Euclidean CFT, the Rényi entropies are obtained from twist operators
\bea
\mathrm{Tr}\rho_A^n\propto\langle{\cal T}_n(u)\bar{{\cal T}}_n(v)\rangle.
\eea
The vacuum Rényi gives the standard scaling:
\bea
\mathrm{Tr}\rho_A^n\sim\bigg(\frac{l}{\epsilon}\bigg)^{-2(\Delta_n+\bar{\Delta}_n)}
=\bigg(\frac{l}{\epsilon}\bigg)^{-\frac{c+\bar{c}}{12}(n-\frac{1}{n})},
\eea
where $l$ is the length of the single interval, $\epsilon$ is the regularization parameter, and the scaling dimensions for the twist and anti-twist operators are:
\bea
\Delta_n=\frac{c}{24}\bigg(n-\frac{1}{n}\bigg); \
\bar{\Delta}_n=\frac{\bar{c}}{24}\bigg(n-\frac{1}{n}\bigg).
\eea
Taking $\partial_n$ at $n=1$ gives:
\bea
S_A=-\partial_n\ln\mathrm{Tr}\rho_A^n\big|_{n=1}=\frac{c+\bar{c}}{6}\ln\bigg(\frac{l}{\epsilon}\bigg)+\mathrm{constant}.
\eea
\\

\noindent
Any local operator ${\cal O}$ with $PT$ invariance supported inside the region $A$ satisfies
\bea
\langle \tilde{0}|{\cal O}|\tilde{0}\rangle=\langle 0| PT{\cal O}PT|0\rangle=\langle 0| {\cal O}|0\rangle,
\eea
which implies locally indistinguishable
\bea
-\mathrm{Tr}\bigg(\rho_A^{(0)}\ln\rho_A^{(0)}\bigg)
\approx-\mathrm{Tr}\bigg(\rho_A^{(\tilde{0})}\ln\rho_A^{(\tilde{0})}\bigg).
\eea
In the holographic limit, we have locality in the CFT.
In any local QFT, the spontaneous breaking of a discrete symmetry implies that distinct vacua become superselected for local operators in the thermodynamical limit
\bea
\langle\tilde{0}|{\cal O}|0\rangle\longrightarrow 0.
\eea
Hence, we get the reduced density matrix for the single interval
\bea
\rho_A\approx \frac{1}{2}\rho_A^{(0)}+\frac{1}{2}\rho_A^{(\tilde{0})}.
\eea
\\

\noindent
The $PT$ transformation exchanges the twist and anti-twist operators, as well as the endpoints of the interval.
Therefore, the twist-pair correlator $\langle{\cal T}_n(u)\bar{{\cal T}}_n(v)\rangle$ for a $PT$-invarant state is $PT$-even, which also implies
\bea
\bar{c}=c^*.
\eea
Hence, we obtain the entanglement entropy
\bea
S_A\approx\frac{c+c^*}{6}\ln\bigg(\frac{l}{\epsilon}\bigg)+\ln 2.
\eea
The superposition's coefficient gives $\ln 2$.
We interpret the coefficient as the maximally mixed between two CFT$_2$ living in the past and future infinity under the holographic limit.
The  central charge of CFT$_2$ for the AdS$_3$ pure Eistein gravity case \cite{Cotler:2018zff,Giombi:2008vd,Huang:2019nfm,Huang:2020tjl} with the analytical continuation
\bea
L\longrightarrow iL
\eea
shows the central charge for the dS$_3$ case
\bea
c=i\frac{3L}{2G_3}+13.
\eea
Hence, we obtain the entanglement entropy
\bea
S_A\approx\frac{13}{3}\ln\bigg(\frac{l}{\epsilon}\bigg)+\ln 2.
\eea
Entanglement entropy only detects the real part of the central charge.
In other words, we lose the classical gravity term for the $ PT$-invariant state.

\subsection{Geodesic}
\noindent
The holographic entanglement entropy for the AdS$_3$ metric is given by \cite{Ryu:2006bv,Ryu:2006ef}
\bea
\frac{l_{\gamma}}{4G_{3, p}}=\frac{3L}{2G_{3, p}}\frac{1}{3}\ln\frac{l}{\epsilon}+\cdots,
\eea
where $l_{\gamma}$ is the length of a geodesic, and $G_{3, p}$ is the 3D physical gravitational constant, which can be identified as the central charge \cite{Cotler:2018zff,Giombi:2008vd,Huang:2019nfm,Huang:2020tjl}:
\bea
c=\frac{3L}{2G_{3, p}}=\frac{3L}{2G_3}+13.
\eea
The $\epsilon$ is the UV cutoff.
Hence, we make the analytical continuation for the holographic entanglement entropy
\bea
\frac{l_{\gamma}}{4G_{3, p}}=\bigg(i\frac{3L}{2G_3}+13\bigg) \frac{1}{3}\ln\frac{l}{\epsilon}+\cdots,
\eea
and its $PT$-pair (after the $PT$-transformation) is
\bea
\frac{l^{PT}_{\gamma}}{4G^*_{3, p}}
=\bigg(-i\frac{3L}{2G_3}+13\bigg) \frac{1}{3}\ln\frac{l}{\epsilon}+\cdots.
\eea
We combine and average two terms to get the logarithmic term for the CFT$_2$ entanglement entropy for a single interval
\bea
\frac{1}{2}\frac{l_{\gamma}}{4G_{3, p}}+\frac{1}{2}\frac{l^{PT}_{\gamma}}{4G_{3, p}}=\frac{13}{3}\ln\frac{l}{\epsilon}+\cdots.
\label{PTP}
\eea
Because the twist-pair correlator is $PT$-invariant, the entanglement entropy is given by the analytical continuation of one and its $PT$-pair.
Therefore, the corresponding holographic entanglement entropy can be made through a similar combination.
Hence, the loss of the classical gravity term is again obtained through holographic entanglement entropy.

\section{Geodesic-Integrated Wightman Function}
\label{sec:6}
\noindent
We construct the timelike geodesic in a manner consistent with the analytical continuation.
Because the detection of the imaginary part of the central charge is necessary to use the single geodesic, we apply it to the geodesic-integrated Wightman function.
Because the dual of the classical gravity term can be done through the timelike geodesic, we expect that the geodesic-integrated Wightman function should be dual to the correlators between a bulk operator and a linear combination of an OPE block \cite{Czech:2015qta,Czech:2016xec,deBoer:2016pqk} and its partner because it is dual to the conformal block for the AdS case.
We derive some useful identities in Appendix \ref{sec:A}.

\subsection{Timelike Geodesic}
\noindent
In the dS case, there is no spacelike geodesic connecting two boundary points $P_1$ and $P_2$.
Instead, the timelike geodesic connects $P_1$ to the antipodal point of $P_2$.
Thus, the dS geodesic becomes
\bea
Y(\lambda)=L\frac{e^{-\lambda}P_1+e^{\lambda}P_{2, A}}{\sqrt{2P_1\cdot P_{2, A}}},
\eea
where
\bea
P_{2, A}=-P_2
\eea
is the antipodal point of $P_2$.
The $Y(\lambda)$ satisfies
\bea
Y^2(\lambda)=1
\eea
so that this curve lies on the dS hyperboloid.
The two asymptotic ends are obtained by sending $\lambda\rightarrow-\infty$ and $\lambda\rightarrow\infty$.
\\

\noindent
For $\lambda\rightarrow-\infty$, the curve approaches the boundary ray $P_1$.
Near the boundary, an early-time cutoff $1/\epsilon$ means that we stop when the coefficient is
\bea
\frac{e^{-\lambda_{\mathrm{mim}}}}{N}=\frac{1}{\epsilon},
\eea
where
\bea
N=\sqrt{2P_1\cdot P_{2, A}}=\sqrt{-2P_1\cdot P_2}.
\eea
Therefore, we get
\bea
\lambda_{\mathrm{min}}=-\ln\frac{N}{\epsilon}.
\eea
For $\lambda\rightarrow\infty$, this approaches the antipodal boundary ray.
The late-time cutoff condition is
\bea
\frac{e^{\lambda_{\mathrm{max}}}}{N}=\frac{1}{\epsilon}.
\eea
Therefore, we get
\bea
\lambda_{\mathrm{max}}=\ln\frac{N}{\epsilon}.
\eea
Because the metric is
\bea
ds^2_{\gamma}=\bigg(\frac{dY}{d\lambda}\bigg)^2d\lambda^2,
\eea
we can combine the two endpoints
\bea
\lambda_{\mathrm{max}}-\lambda_{\mathrm{min}}
=2\ln\frac{N}{\epsilon}
\eea
to obtain the regulated proper length of the timelike geodesic
\bea
l_{\gamma}=2iL\ln\frac{N}{\epsilon}.
\eea
Hence, we can use the $PT$-pair of geodesics to extract the real part of the central charge with the same result as in Eq. \ref{PTP}.
To detect the imaginary part of the central charge, we need to use the single geodesic
\bea
\frac{l_{\gamma}}{4G_{3, p}}
\sim\frac{iL}{2G_3}\ln\frac{\sqrt{-2P_1\cdot P_2}}{\epsilon}
=i\frac{\mathrm{Im}(c)}{3} \ln\frac{\sqrt{-2P_1\cdot P_2}}{\epsilon}.
\eea
The imaginary part of the central charge corresponds to
\bea
\mathrm{Im}(c)=\frac{3L}{2G_3}.
\eea
We can use the boundary-space embedding vector
\bea
P(x)=(P^+, P^-, P^{\mu})=(1, t^2-\vec{x}\cdot\vec{x}, x^{\mu}),
\eea
where
\bea
x^{\mu}=(t, \vec{x}).
\eea
The embedding inner product is
\bea
P_1\cdot P_2=-\frac{1}{2}(P_1^+P_2^-+P_1^-P_2^+)
+\eta_{\mu\nu}P_1^{\mu}P_2^{\mu},
\eea
where $\eta_{\mu\nu}=(1, -1, -1, \cdots, -1)$.
Therefore, we get
\bea
-2P_1\cdot P_2=(t_1-t_2)^2-(\vec{x}_1-\vec{x}_2)\cdot(\vec{x}_1-\vec{x}_2)\equiv
(\Delta t)^2-(\Delta x)^2.
\eea
Hence, the $\sqrt{-2P_1\cdot P_2}$ is the timelike interval length, and the holographic entanglement entropy formula becomes
\bea
\frac{l_{\gamma}}{4G_{3, p}}
\sim
i\frac{\mathrm{Im}(c)}{3} \ln\frac{\sqrt{-2P_1\cdot P_2}}{\epsilon}
=i\frac{\mathrm{Im}(c)}{3} \ln\frac{\sqrt{(\Delta t)^2-(\Delta x)^2}}{\epsilon}.
\eea
We can now use the holographic entanglement entropy formula \cite{Ryu:2006bv,Ryu:2006ef} to detect the classical gravity term via a timelike geodesic.
Hence, we expect that the Wightman function along the single geodesic integration will yield a result similar to that in the AdS case.
\\

\noindent
We would also like to point out that, geometrically, such a result of a timelike geodesic can be obtained from the discontinuous geodesics used in the computation of pseudo entropy \cite{Nakata:2020luh,Doi:2022iyj} in dS space (of the single interval specified by $P_{1,2}$).
For the holographic pseudo entropy in the dS case, there are two portions of timelike geodesics connected by a spacelike one, whose length is an independent constant.
The antipodal map of the portion, together with $P_2$, leads to the smooth timelike geodesic we consider here.
The timelike geodesic result captures the logarithmic term in the pseudo-entropy.
The pseudo-entropy is the entanglement entropy with a reduced transition matrix for two states, which differs from the reduced density matrix, which only uses one state \cite{Nakata:2020luh}.

\subsection{Geodesic Integration}
\noindent
The Wightman function for a real scalar field in the Bunch-Davies vacuum is \cite{Bousso:2001mw}
\bea
G_E(x, y)=C_{\Delta}{}_2F_1\bigg(\Delta_+, \Delta_-; \frac{d+1}{2}; \frac{1+Z(x, y)}{2}\bigg),
\eea
where
\bea
C_{\Delta}\equiv\frac{\Gamma(\Delta_+)\Gamma(\Delta_-)}{(4\pi)^{\frac{d+1}{2}}\Gamma\big(\frac{d+1}{2}\big)}.
\eea
We let
\bea
Z(x, y)=\cos\big(D_{\mathrm{dS}}(x, y)\big).
\eea
We then have
\bea
u(x, y)\equiv\cos^2\bigg(\frac{D_{\mathrm{dS}}(x, y)}{2}\bigg)=\frac{1+Z(x, y)}{2}.
\eea
If we now assume that $y=Y(\lambda)$ is a timelike geodesic parametrized by proper time $\lambda$, the geodesic Wightman function is simply
\bea
I_{\gamma}(x)=\int_{\gamma}d\lambda\ G_E\big(x, Y(\lambda)\big),
\eea
or explicitly
\bea
I_{\gamma}(x)=C_{\Delta}\int d\lambda\ {}_2F_1\bigg(\Delta_+, \Delta_-; \frac{d+1}{2}; \frac{1+Z(x, \lambda)}{2}\bigg),
\eea
where
\bea
Z\big(x, \lambda\big)=\cos D_{\mathrm{dS}}\big(x, Y(\lambda)\big).
\eea
We use the timelike geodesic parametrization
\bea
Y(\lambda)=\frac{e^{-\lambda}P_1-e^{\lambda}P_2}{\sqrt{-2P_1\cdot P_2}}.
\eea
We then define
\bea
N\equiv\sqrt{-2P_1\cdot P_2}; \
\alpha\equiv\frac{P_1\cdot X}{N}; \
\beta\equiv -\frac{P_2\cdot X}{N},
\eea
and then obtain
\bea
Y(\lambda)\cdot X=\alpha e^{-\lambda}+\beta e^{\lambda}.
\eea
Therefore, the geodesic-integrated Bunch-Davies Wightman function is
\bea
I_{\gamma}(X)&=&C_{\Delta}\int d\lambda\ {}_2F_1\bigg(\Delta_+, \Delta_-; \frac{d+1}{2}; \frac{1+Y(\lambda)\cdot X}{2}\bigg)
\nn\\
&=&C_{\Delta}\int d\lambda\ {}_2F_1\bigg(\Delta_+, \Delta_-; \frac{d+1}{2}; \frac{1+\alpha e^{-\lambda}+\beta e^{\lambda}}{2}\bigg).
\eea
\\

\noindent
If the full time-like geodesic is integrated, $\lambda\in (-\infty, \infty)$, then one can shift
\bea
\lambda=\tilde{\lambda}+\lambda_0; \ \lambda_0=\frac{1}{2}\ln\frac{\alpha}{\beta}
\eea
when $\alpha\beta>0$.
We also have
\bea
\alpha e^{-\lambda}+\beta e^{\lambda}\rightarrow 2\sqrt{\alpha\beta}\cosh(\lambda).
\eea
Therefore, we get
\bea
I_{\gamma}(X)=C_{\Delta}\int^{\infty}_{-\infty}d\lambda\ {}_2F_1\bigg(\Delta_+, \Delta_-; \frac{d+1}{2};
\frac{1+2\sqrt{\alpha\beta}\cosh\lambda}{2}\bigg).
\eea
For $\alpha\beta<0$, instead,
\bea
\alpha e^{-\lambda}+\beta e^{\lambda}=\mp 2\sqrt{-\alpha\beta}\sinh(\lambda-\lambda_0),
\eea
where
\bea
\lambda_0=\frac{1}{2}\ln\bigg(-\frac{\alpha}{\beta}\bigg).
\eea
The minus sign corresponds to
\bea
\alpha>0; \ \beta<0.
\eea
For the plus sign, it corresponds to
\bea
\alpha<0; \ \beta>0.
\eea
Therefore, the result becomes
\bea
I_{\gamma}(X)=C_{\Delta}\int^{\infty}_{-\infty}d\lambda\ {}_2F_1\bigg(\Delta_+, \Delta_-; \frac{d+1}{2};
\frac{1-2\eta \sqrt{-\alpha\beta}\sinh(\lambda)}{2}\bigg),
\eea
where
\bea
\eta\equiv\pm 1.
\eea
The form of $\cosh$ or $\sinh$ depends on the sign of $\alpha\beta$.
Therefore, we consider two cases ($\alpha\beta>0$ and $\alpha\beta<0$) to discuss the result of $I_{\gamma}$.
\\

\noindent
We now set
\bea
L=1
\eea
for simplicity and to analyze the value of
\bea
q\equiv 2\sqrt{|\alpha\beta|}
\eea
by introducing the basis
\bea
n_0=\frac{P_1-P_2}{\sqrt{-2P_1\cdot P_2}}; \
n_1=\frac{-P_2-P_1}{\sqrt{-2P_1\cdot P_2}}.
\eea
We then have
\bea
n_0^2=1; \ n_1^2=-1; \ n_0\cdot n_1=0.
\eea
The dS timelike geodesic is
\bea
Y(\lambda)=n_0\cosh(\lambda)+n_1\sinh(\lambda).
\eea
The bulk point is decomposed as
\bea
X=A_1 n_0+ A_2 n_1+X_{\perp},
\eea
where
\bea
X_{\perp}\cdot n_0=X_{\perp}\cdot n_1=0.
\eea
Because we have
\bea
X^2=1,
\eea
we obtain
\bea
A_1^2-A_2^2+X_{\perp}^2=1.
\eea
We also have
\bea
Y(\lambda)\cdot X=A_1\cosh(\lambda)-A_2\sinh(\lambda).
\eea
When $\alpha\beta>0$, the parameters are
\bea
A_1=q>0; \ A_2=0.
\eea
Therefore, we get
\bea
q^2=1-X_{\perp}^2>0.
\eea
In other words, we get
\bea
0<q\le 1.
\eea
When $\alpha\beta<0$,we choose the parameters
\bea
A_1=0; \ A_2=\eta q.
\eea
Therefore, we get
\bea
q^2=X_{\perp}^2-1.
\eea
In other words, the $q$ does not have an upper bound.
\\

\noindent
We can define the timelike geodesic by
\bea
Y(\lambda)=n_0\cosh(\lambda)+n_1\sinh(\lambda).
\eea
To have
\bea
Y(\lambda)\cdot X\sim q\cosh(\lambda)
\eea
up to a shift of $\lambda$ by a real-valued constant for $\alpha\beta>0$, the dS bulk point is decomposed as
\bea
X=q\big(n_0\cosh(t)+n_1\sinh(t)\big)+\sqrt{1-q^2}\Omega_{d-1},
\eea
where
\bea
\Omega_{d-1}^2=1, \ \Omega_{d-1}\cdot n_0=\Omega_{d-1}\cdot n_1=0.
\eea
We then obtain
\bea
dX^2=\frac{dq^2}{1-q^2}-q^2dt^2+(1-q^2)d\Omega_{d-1}^2.
\eea
Equivalently, we use the following identifications
\bea
Y(\lambda)\cdot X=\cos\bigg(D_{\mathrm{dS}}\big(X, Y(\lambda)\big)\bigg)=\cos(\theta)\cosh(\lambda); \ q=\cos(\theta)
\eea
to obtian the dS metric
\bea
ds^2=d\theta^2-\cos^2(\theta)dt^2+\sin^2(\theta)d\Omega_{d-1}^2.
\label{sp}
\eea
\\

\noindent
For $\alpha\beta<0$, we need to have
\bea
Y(\lambda)\cdot X\sim -\eta q\sinh(\lambda)
\eea
up to a shift of $\lambda$ by a real-valued constant, the dS bulk point becomes
\bea
X=-\eta q\big(n_0\sinh(t)+n_1\cosh(t)\big)+\sqrt{1+q^2}\Omega_{d-1}.
\eea
The metric is
\bea
dX^2=-\frac{dq^2}{1+q^2}+q^2dt^2+(1+q^2)d\Omega_{d-1}^2.
\eea
We then apply
\bea
q=\sinh(\theta)
\eea
to obtain the metric
\bea
ds^2=-d\theta^2+\sinh^2(\theta)dt^2+\cosh^2(\theta)d\Omega_{d-1}^2.
\label{cosh}
\eea
\\

\noindent
When $\alpha\beta>0$, it corresponds to
\bea
X_{\perp}\le 1.
\eea
This is the static-patch regime.
If $0<\theta\le\pi/2$, it is one static diamond.
If $\pi/2\le\theta<\pi$, it describes the antipodal static diamond.
The surface for $\theta=\pi/2$ is the cosmological horizon.
When $\alpha\beta<0$, it corresponds to
\bea
X_{\perp}\ge 1.
\eea
This describes the cosmological regions outside the static horizons.
Hence, Eqs. \eqref{sp} and \eqref{cosh} do not form one smooth coordinate chart, but they are glued across the horizon $X_{\perp}=1$ to cover a global dS patchwise.

\subsubsection{$\alpha\beta>0$}
\noindent
We compute the geodesic-integrated Wightman function for $\alpha\beta>0$,
\bea
I_{\gamma}(X)=C_{\Delta}\int^{\infty}_{-\infty}d\lambda\ {}_2F_1\bigg(\Delta_+, \Delta_-; \frac{d+1}{2}; \frac{1+q\cosh\lambda}{2}\bigg),
\eea
through the Barnes representation of the hypergeometric function
\bea
{}_2F_1(A_1, A_2; A_3; z)=
\frac{\Gamma(A_3)}{\Gamma(A_1)\Gamma(A_2)}
\int^{i\infty}_{-i\infty}\frac{ds}{2\pi i}\
\frac{\Gamma(A_1+s)\Gamma(A_2+s)\Gamma(-s)}
{\Gamma(A_3+s)}(-z)^s.
\eea
Because we have
\bea
C_{\Delta}\frac{\Gamma\big(\frac{d+1}{2}\big)}{\Gamma(\Delta_+)\Gamma(\Delta_-)}
=\frac{1}{(4\pi)^{\frac{d+1}{2}}},
\eea
we obtain
\bea
&&
I_{\gamma}(X)
\nn\\
&=&\frac{1}{(4\pi)^{\frac{d+1}{2}}}
\int_{\sigma-i\infty}^{\sigma+i\infty}\frac{ds}{2\pi i}\
\frac{\Gamma(\Delta_++s)\Gamma(\Delta_-+s)\Gamma(-s)}{\Gamma\big(\frac{d+1}{2}+s\big)}
\int^{\infty}_{-\infty}d\lambda\ \bigg(-\frac{1+q\cosh\lambda}{2}\bigg)^s.
\nn\\
\eea
To have the convergent integration, we shift the integration range by $\sigma$ to have the negative real part for $s$, where
\bea
-\min\big(\mathrm{Re}(\Delta_+), \mathrm{Re}(\Delta_-)\big)<\sigma<0.
\eea
The hypergeometric function for the Barnes representation is unaffected by the shift of $s$.
\\

\begin{figure}[tbp]
\centering
\includegraphics[scale = 0.77]{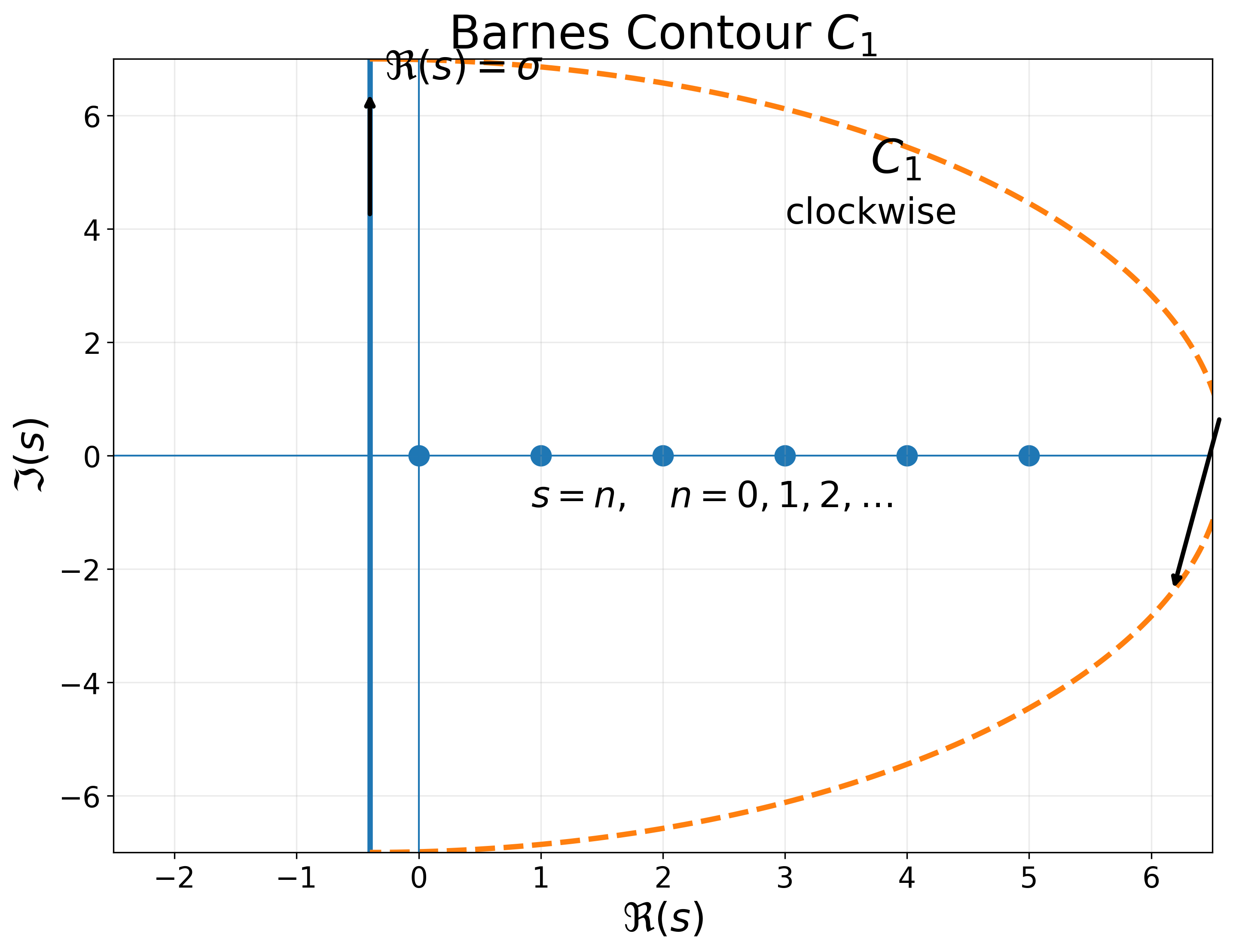}
\caption{The Barnes clockwise contour $C_1$ encloses the poles at $s=n$, where $n=0, 1, \cdots$.}
\label{C1}
\end{figure}

\noindent
We now compute the relevant integration
\bea
I_s(A, B)=\int_{-\infty}^{\infty}d\lambda\ (A+B\cosh\lambda)^s
\eea
with $\mathrm{Re}(s)<0$.
Because the integrand is even for $\lambda$, we can get
\bea
I_s(A, B)=2\int^{\infty}_0d\lambda\ (A+B\cosh\lambda)^s.
\eea
We then use
\bea
u=\tanh^2\bigg(\frac{\lambda}{2}\bigg),
\eea
to obtain
\bea
\cosh(\lambda)=\frac{1+u}{1-u}; \ d\lambda=\frac{du}{\sqrt{u}(1-u)}.
\eea
Because the range of $\lambda$ is $\lambda\in\lbrack 0, \infty)$, we have
$u\in\lbrack 0, 1)$.
Hence, we obtain
\bea
I_s&=&2\int_0^{1}\frac{du}{\sqrt{u}(1-u)}\ \bigg(A+B\frac{1+u}{1-u}\bigg)
\nn\\
&=&2\int_0^1du\ u^{-\frac{1}{2}}(1-u)^{-1}\bigg(\frac{(A+B)+(B-A)u}{1-u}\bigg)^s
\nn\\
&=&2(A+B)^s\int_0^1du\ u^{-\frac{1}{2}}(1-u)^{-s-1}\bigg(1-\frac{A-B}{A+B}u\bigg)^s.
\eea
We now compare it with Euler's integral representation
\bea
{}_2F_1(a_1, a_2; a_3; x)=\frac{\Gamma(a_3)}{\Gamma(a_2)\Gamma(a_3-a_2)}
\int_0^1du\ u^{a_2-1}(1-u)^{a_3-a_2-1}(1-xu)^{-a_1}
\eea
with
\bea
\mathrm{Re}(a_3)>\mathrm{Re}(a_2)>0
\eea
and that $x$ is not a real number greater than or equal to 1.
Therefore, we select
\bea
a_1=-s; \ a_2=\frac{1}{2}; \ a_3=\frac{1}{2}-s ; \ x=\frac{A-B}{A+B}
\eea
to identify the integration as
\bea
I_s=2(A+B)^s\frac{\Gamma\big(\frac{1}{2}\big)\Gamma(-s)}{\Gamma\big(\frac{1}{2}-s\big)}
{}_2F_1\bigg(-s, \frac{1}{2}; \frac{1}{2}-s; \frac{A-B}{A+B}\bigg).
\eea
Hence, the geodesic-integrated Bunch-Davies Wightman function is
\bea
&&
I_{\gamma}(X)
\nn\\
&=&\frac{2\sqrt{\pi}}{(4\pi)^{\frac{d+1}{2}}}
\nn\\
&&\times
\int^{\sigma+i\infty}_{\sigma-i\infty}\frac{ds}{2\pi i}\
\frac{\Gamma(\Delta_++s)\Gamma(\Delta_-+s)\big(\Gamma(-s)\big)^2}
{\Gamma\big(\frac{d+1}{2}+s\big)\Gamma\big(\frac{1}{2}-s\big)}
\nn\\
&&\times
\bigg(-\frac{1+q}{2}\bigg)^s
{}_2F_1\bigg(-s, \frac{1}{2}; \frac{1}{2}-s; \frac{1-q}{1+q}\bigg).
\eea
We use the right and clockwise contour $C_1$ (Fig. \ref{C1}) enclosing the double poles at
\bea
s=n, \ n=0, 1, \cdots
\eea
to compute $I_{\gamma}$,
\bea
I_{\gamma}(X)=\frac{2\sqrt{\pi}}{(4\pi)^{\frac{d+1}{2}}}\oint_{C_1}\frac{ds}{2\pi i}\
\frac{\Gamma(\Delta_++s)\Gamma(\Delta_-+s)\big(\Gamma(-s)\big)^2}
{\Gamma\big(\frac{d+1}{2}+s\big)\Gamma\big(\frac{1}{2}-s\big)}
Q^s{}_2F_1\bigg(-s, \frac{1}{2}; \frac{1}{2}-s; W\bigg),
\nn\\
\eea
where
\bea
Q\equiv-\frac{1+q}{2}; \
W\equiv\frac{1-q}{1+q}.
\eea

\subsubsection{$\alpha\beta<0$}
\noindent
We now consider $\alpha\beta<0$, and the geodesic-integrated Wightman function is given by
\bea
I_{\gamma}(X)=C_{\Delta}\int^{\infty}_{-\infty}d\lambda\
{}_2F_1\bigg(\Delta_+, \Delta_-; \frac{d+1}{2}; \frac{1-\eta q\sinh\lambda}{2}\bigg).
\eea
We then use the Barnes representation to obtain
\bea
I_{\gamma}(X)=\frac{1}{(4\pi)^c}\int^{\sigma+i\infty}_{\sigma-i\infty}\frac{ds}{2\pi i}\
\frac{\Gamma(\Delta_++s)\Gamma(\Delta_-+s)\Gamma(-s)}{\Gamma\big(\frac{d+1}{2}+s\big)}K_s(q),
\eea
where the $\lambda$-integral is
\bea
K_s(q)=\int^{\infty}_{-\infty}d\lambda\ \bigg(-\frac{1-\eta q\sinh\lambda}{2}\bigg)^s.
\eea
We define a new variable
\bea
x=\lambda+i\frac{\pi}{2},
\eea
which gives
\bea
\sinh(\lambda)=-i\cosh(x).
\eea
Therefore, we get
\bea
K_s(q)&=&\int^{\infty}_{-\infty}dx\ \bigg(-\frac{1+i\eta q\cosh x}{2}\bigg)^s
\nn\\
&=&2\bigg(-\frac{1+i\eta q}{2}\bigg)^s\frac{\Gamma\big(\frac{1}{2}\big)\Gamma(-s)}{\Gamma\big(\frac{1}{2}-s\big)}
{}_2F_1\bigg(-s, \frac{1}{2}; \frac{1}{2}-s; \frac{1-i\eta q}{1+i\eta q}\bigg).
\eea
Before discussing the poles, we can already see that the result is the analytical continuation of the $\alpha\beta>0$ case through
\bea
q\rightarrow i\eta q.
\eea
\\

\begin{figure}[tbp]
\centering
\includegraphics[scale = 0.35]{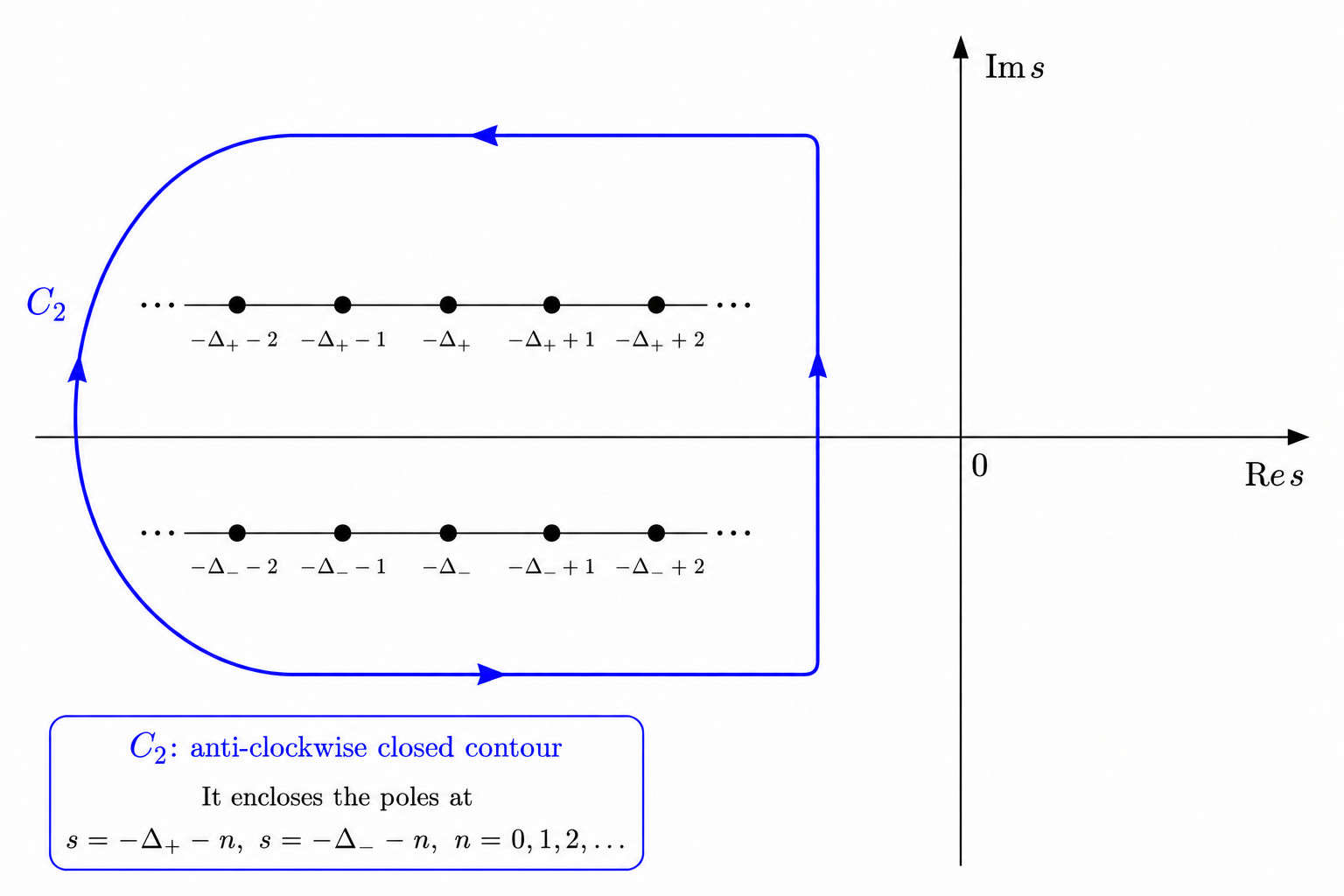}
\caption{The Barnes anti-clockwise contour $C_2$ encloses the poles at $s=-\Delta_{\pm}-n$, where $n=0, 1, \cdots$.}
\label{C2}
\end{figure}

\noindent
When $|q|<\sqrt{3}$, we close the Barnes clockwise contour $C_1$ to the right enclosing the double poles of $\big(\Gamma(-s)\big)^2$,
\bea
s=n, \ n=0, 1, \cdots.
\eea
For the case of $|q|>\sqrt{3}$, we close the Barnes anti-clockwise contour $C_2$ to the left (Fig. \ref{C2}).
The left poles come from $\Gamma(\Delta_++s)$ and $\Gamma(\Delta_-+s)$, namely
\bea
s=-\Delta_+-n, \
s=-\Delta_--n, \
n=0, 1, \cdots.
\eea
We first assume that
\bea
\Delta_+-\Delta_-\notin\mathbb{Z},
\eea
so that the two-pole families do not collide.
If $\Delta_+-\Delta_-\in\mathbb{Z}$, the two pole families collide, and we need to get the result by identifying $\Delta_-=\Delta_++m+\epsilon$, $m=0, 1, \cdots$, and then taking the limit $\epsilon\rightarrow 0$.

\subsection{Hypergeometric Function}
\noindent
The bulk equation of motion for a real scalar field in the dS background is
\bea
\bigg(\nabla_{\mathrm{dS}}^2-m^2\bigg)I_{\gamma}(X)=\delta_{\gamma}(X),
\label{dseom}
\eea
where
\bea
m^2L^2=\Delta(d-\Delta)=\Delta_+\Delta_-.
\eea
We now set $L=1$ for simplicity and also define
\bea
u\equiv\sin^2(\theta),
\eea
and then the equation \eqref{dseom} awaying from the geodesic becomes
\bea
u(1-u)F^{\prime\prime}(u)
+\bigg\lbrack\frac{d}{2}-\bigg(\frac{d}{2}+1\bigg)u\bigg\rbrack F^{\prime}(u)
-\frac{\Delta(d-\Delta)}{4}F(u)=0,
\eea
where
\bea
I_{\gamma}(X)=F(u).
\eea
This is the hypergeometric equation
\bea
u(1-u)F^{\prime\prime}(u)
+\big(c-(a+b+1)u\big) F^{\prime}(u)
-abF(u)=0
\eea
with the following choice of the parameters:
\bea
a=\frac{\Delta}{2}=\frac{\Delta_+}{2}; \
b=\frac{d-\Delta}{2}=\frac{\Delta_-}{2}; \
c=\frac{d}{2}=\frac{\Delta_++\Delta_-}{2}.
\eea
Therefore, the general solution is
\bea
F(u)=C_1 \cdot {}_2F_1\bigg(\frac{\Delta}{2}, \frac{d-\Delta}{2}; \frac{d}{2}; u\bigg)
+C_2u^{1-\frac{d}{2}}{}_2F_1\bigg(1-\frac{\Delta}{2}, 1-\frac{d-\Delta}{2}; 2-\frac{d}{2}; u\bigg).
\label{gshg}
\eea
Returing to $\theta$, we get
\bea
&&
I_{\gamma}(\rho)
\nn\\
&=&C_1 \cdot {}_2F_1\bigg(\frac{\Delta}{2}, \frac{d-\Delta}{2}; \frac{d}{2}; \sin^2\theta\bigg)
\nn\\
&&
+C_2(\sin^2\theta)^{1-\frac{d}{2}}{}_2F_1\bigg(1-\frac{\Delta}{2}, 1-\frac{d-\Delta}{2}; 2-\frac{d}{2}; \sin^2\theta\bigg).
\eea
\\

\noindent
We first consider $\alpha\beta>0$ and then the scalar Laplacian becomes
\bea
\nabla^2 f(q)=\frac{1}{q(1-q^2)^{\frac{d-2}{2}}}\partial_q\big(q(1-q^2)^{\frac{d}{2}}\partial_qf\big)
=(1-q^2)f^{\prime\prime}(q)+\frac{1-(d+1)q^2}{q}f^{\prime}(q).
\eea
Therefore, the geodesic-integrated object satisfies
\bea
\bigg((1-q^2)\partial_q^2+\frac{1-(d+1)q^2}{q}\partial_q-m^2\bigg)I_{\gamma}(q)\sim \delta(q-1),
\eea
where the timelike geodesic line corresponds to $q=1$.
Since we have
\bea
m^2=\Delta_+\Delta_-,
\eea
the operator corresponding to the hypergeometric equation is
\bea
{\cal L}_q=(1-q^2)\partial_q^2+\frac{1-(d+1)q^2}{q}\partial_q-\Delta_+\Delta_-.
\eea
\\

\noindent
We define the term having the $q$-dependence in the $I_{\gamma}(X)$,
\bea
P_s(q)&\equiv&\bigg(-\frac{1+q}{2}\bigg)^s
{}_2F_1\bigg(-s, \frac{1}{2}; \frac{1}{2}-s; \frac{1-q}{1+q}\bigg)
=(-1)^s(1+x)^{-s}y(x)
\nn\\
&\equiv&(-1)^s P,
\eea
where
\bea
x\equiv\frac{1-q}{1+q}; \
y(x)\equiv{}_2F_1\bigg(-s, \frac{1}{2}; \frac{1}{2}-s; x\bigg).
\eea
Because $y(x)$ is the hypergeometric function, it satisfies
\bea
x(1-x)y^{\prime\prime}
+\bigg\lbrack\frac{1}{2}-s-\bigg(\frac{3}{2}-s\bigg)x\bigg\rbrack y^{\prime}
+\frac{s}{2}y=0.
\eea
We now compute the $q$-derivative
\bea
\frac{dx}{dq}=-\frac{2}{(1+q)^2}=-\frac{(1+x)^2}{2}.
\eea
Thus, we get
\bea
\partial_q=-\frac{(1+x)^2}{2}\partial_x.
\eea
We also have
\bea
1-q^2=1-\bigg(\frac{1-x}{1+x}\bigg)^2=\frac{4x}{(1+x)^2}.
\eea
Therefore, a direct computation gives
\bea
&&
(1-q^2)P_{qq}+\frac{1+(2s-2)q^2}{q}P_q-s(s-1)P
\nn\\
&=&
\frac{(1+x)^2}{1-x}(1+x)^{-s}
\Bigg\lbrack x(1-x)y^{\prime\prime}
+\bigg\lbrack\frac{1}{2}-s-\bigg(\frac{3}{2}-s\bigg)x \bigg\rbrack y^{\prime}
+\frac{s}{2}y\Bigg\rbrack.
\eea
Because the $y$ satisfies the hypergeometric equation
\bea
x(1-x)y^{\prime\prime}
+\bigg\lbrack\frac{1}{2}-s-\bigg(\frac{3}{2}-s\bigg)x\bigg\rbrack y^{\prime}
+\frac{s}{2}y=0,
\eea
we get
\bea
(1-q^2)P_s^{\prime\prime}+\frac{1+(2s-2)q^2}{q}P_s^{\prime}-s(s-1)P_s=0.
\eea
Therefore, we get
\bea
{\cal L}_qP_s=-(2s+d-1)qP_s^{\prime}+\big(s(s-1)-\Delta_+\Delta_-\big)P_s.
\eea
In Appendix \ref{sec:A}, we prove the following relation
\bea
&&
(1-x){}_2F_1\bigg(1-s, \frac{3}{2}; \frac{3}{2}-s; x\bigg)
\nn\\
&=&-(2s-1){}_2F_1\bigg(-s, \frac{1}{2}; \frac{1}{2}-s; x\bigg)
+2s{}_2F_1\bigg(1-s, \frac{1}{2}; \frac{3}{2}-s; x\bigg)
\eea
and use it to prove the following identity
\bea
qP_s^{\prime}=sP_s-\frac{s^2}{1-2s}P_{s-1}.
\eea
Hence, we obtain
\bea
{\cal L}_qP_s=-(s+\Delta_+)(s+\Delta_-)P_s
+\frac{(2s+d-1)s^2}{1-2s}P_{s-1}.
\eea
\\

\noindent
The Barnes integral for $\alpha\beta>0$ and $q<1$ becomes
\bea
I_{\gamma}=\oint_{C_1}\frac{ds}{2\pi i}\ C(s)P_s(q),
\eea
where
\bea
C(s)\equiv\frac{2\sqrt{\pi}}{(4\pi)^{\frac{d+1}{2}}}\frac{\Gamma(\Delta_++s)\Gamma(\Delta_-+s)\big(\Gamma(-s)\big)^2}
{\Gamma\big(\frac{d+1}{2}+s\big)\Gamma\big(\frac{1}{2}-s\big)}.
\eea
Therefore, we get
\bea
{\cal L}_qI_{\gamma}&=&\oint_{C_1}\frac{ds}{2\pi i}\ C(s){\cal L}_qP_s(q)
\nn\\
&=&\oint_{C_1}\frac{ds}{2\pi i}\ C(s)\bigg(-(s+\Delta_+)(s+\Delta_-)P_s
+\frac{(2s+d-1)s^2}{1-2s}P_{s-1}\bigg).
\eea
In Appendix \ref{sec:A}, we show that the $C(s)$ satisfies the following relation
\bea
C(s)\frac{(2s+d-1)s^2}{1-2s}P_{s-1}
=C(s-1)(s+\Delta_+-1)(s+\Delta_--1)P_{s-1}.
\eea
We then make the shifting of $s$ as
\bea
C(s-1)(s+\Delta_+-1)(s+\Delta_--1)P_{s-1}\rightarrow C(s)(s+\Delta_+)(s+\Delta_-)P_{s}
\eea
because it does include the poles at $s=0$ for $\alpha\beta>0$.
Therefore, we get
\bea
{\cal L}_qI_{\gamma}=0, \ q<1.
\eea
The above result also holds for $\alpha\beta<0$ and $q<\sqrt{3}$ because we just need to replace $q$ with $i\eta q$.
When $\alpha\beta<0$ and $q>\sqrt{3}$, we consider the left poles, but we do not have the poles at $s=-\Delta_{\pm}$ in $(s+\Delta_+)(s+\Delta_-)P_s$, so that we can shift $s$ to reach the same conclusion
\bea
{\cal L}_qI_{\gamma}=0.
\eea
Hence, we conclude that the geodesic-integrated Bunch-Davies Wightman function satisfies the hypergeometric equation for the global dS.

\subsection{OPE Block and its Partner}
\noindent
We start to write the general solution of $I_{\gamma}$ \eqref{gshg} in terms of the positive cross ratio
\bea
\chi\equiv\frac{4|(P_1\cdot X)(P_2\cdot X)|}{-2(P_1\cdot P_2)}=4|\alpha\beta|=q^2=\cos^2(\theta).
\eea
The general solution can be rewritten as
\bea
&&
I_{\gamma}(\rho)
\nn\\
&=&C_1 \cdot {}_2F_1\bigg(a, b; c; 1-\chi\bigg)
\nn\\
&&
+C_2(1-\chi)^{1-c}{}_2F_1\bigg(1+a-c, 1+b-c; 2-c; 1-\chi\bigg)
\nn\\
&=&
C_1 \cdot {}_2F_1\bigg(a, b; a+b; 1-\chi\bigg)
\nn\\
&&
+C_2(1-\chi)^{1-a-b}{}_2F_1\bigg(1-b, 1-a; 2-a-b; 1-\chi\bigg),
\eea
where
\bea
a=\frac{\Delta_+}{2}; \
b=\frac{\Delta_-}{2}; \
c=a+b=\frac{d}{2}.
\eea
We now want to use the hypergeometric connection formula,
\bea
&&
{}_2F_1(a, b; c; z)
\nn\\
&=&A_1(-z)^{-a}{}_2F_1\bigg(a, 1+a-c; 1+a-b; \frac{1}{z}\bigg)
\nn\\
&&
+A_2(-z)^{-b}{}_2F_1\bigg(b, 1+b-c; 1+b-a; \frac{1}{z}\bigg),
\eea
where
\bea
A_1=\frac{\Gamma(c)\Gamma(b-a)}{\Gamma(b)\Gamma(c-a)}; \
A_2=\frac{\Gamma(c)\Gamma(a-b)}{\Gamma(a)\Gamma(c-b)}.
\eea
We then choose
\bea
z=1-\chi; \
c=a+b
\eea
and use
\bea
c-a=b; \ c-b=a
\eea
and
\bea
1+a-c=1-b; \ 1+b-c=1-a
\eea
to show
\bea
&&
{}_2F_1(a, b; a+b; 1-\chi)
\nn\\
&=&
\frac{\Gamma(a+b)\Gamma(b-a)}{\big(\Gamma(b)\big)^2}(\chi-1)^{-a}{}_2F_1\bigg(a, 1-b; 1+a-b; \frac{1}{1-\chi}\bigg)
\nn\\
&&+
\frac{\Gamma(a+b)\Gamma(a-b)}{\big(\Gamma(a)\big)^2}(\chi-1)^{-b}{}_2F_1\bigg(b, 1-a; 1+b-a; \frac{1}{1-\chi}\bigg).
\eea
\\

\noindent
Using Pfaff's transformation
\bea
{}_2F_1(A, B; C; x)=(1-x)^{-A}{}_2F_1\bigg(A, C-B; C; \frac{x}{x-1}\bigg)
\eea
with
\bea
x=\frac{1}{1-\chi}; \
\frac{x}{x-1}=\frac{1}{\chi}; \
1-x=\frac{\chi}{\chi-1},
\eea
we then get
\bea
(\chi-1)^{-a}{}_2F_1\bigg(a, 1-b; 1+a-b; \frac{1}{1-\chi}\bigg)
=\chi^{-a}{}_2F_1\bigg(a, a; 1+a-b; \frac{1}{\chi}\bigg)
\eea
and
\bea
(\chi-1)^{-b}{}_2F_1\bigg(b, 1-a; 1+b-a; \frac{1}{1-\chi}\bigg)
=\chi^{-b}{}_2F_1\bigg(b, b; 1+b-a; \frac{1}{\chi}\bigg).
\eea
Hence, we obtain
\bea
&&
{}_2F_1(a, b; a+b; 1-\chi)
\nn\\
&=&\frac{\Gamma(a+b)\Gamma(b-a)}{\big(\Gamma(b)\big)^2}\chi^{-a}
{}_2F_1\bigg(a, a; 1+a-b; \frac{1}{\chi}\bigg)
\nn\\
&&+
\frac{\Gamma(a+b)\Gamma(a-b)}{\big(\Gamma(a)\big)^2}\chi^{-a}
{}_2F_1\bigg(b, b; 1+b-a; \frac{1}{\chi}\bigg).
\eea
We now want to apply the hypergeometric connection formula to get another identity
\bea
&&
{}_2F_1(1-b, a; 1+a-b; 2-a-b; 1-\chi)
\nn\\
&=&
\frac{\Gamma(2-a-b)\Gamma(b-a)}{\big(\Gamma(1-a)\big)^2}
(\chi-1)^{b-1}{}_2F_1\bigg(1-b, a; 1+a-b; \frac{1}{1-\chi}\bigg)
\nn\\
&&+
\frac{\Gamma(2-a-b)\Gamma(a-b)}{\big(\Gamma(1-b)\big)^2}
(\chi-1)^{a-1}{}_2F_1\bigg(1-a, b; 1+b-a; \frac{1}{1-\chi}\bigg)
\nn\\
&=&
\frac{\Gamma(2-a-b)\Gamma(b-a)}{\big(\Gamma(1-a)\big)^2}
(\chi-1)^{b-1}{}_2F_1\bigg(a, 1-b; 1+a-b; \frac{1}{1-\chi}\bigg)
\nn\\
&&+
\frac{\Gamma(2-a-b)\Gamma(a-b)}{\big(\Gamma(1-b)\big)^2}
(\chi-1)^{a-1}{}_2F_1\bigg(b, 1-a; 1+b-a; \frac{1}{1-\chi}\bigg).
\eea
By multiplying the prefactor
\bea
(1-\chi)^{1-a-b}=(-1)^{1-a-b}(\chi-1)^{1-a-b},
\eea
we obtain
\bea
&&
(1-\chi)^{1-a-b}{}_2F_1(1-b, a; 1+a-b; 2-a-b; 1-\chi)
\nn\\
&=&
(-1)^{1-a-b}\frac{\Gamma(2-a-b)\Gamma(b-a)}{\big(\Gamma(1-a)\big)^2}
(\chi-1)^{-a}{}_2F_1\bigg(a, 1-b; 1+a-b; \frac{1}{1-\chi}\bigg)
\nn\\
&&+
(-1)^{1-a-b}
\frac{\Gamma(2-a-b)\Gamma(a-b)}{\big(\Gamma(1-b)\big)^2}
(\chi-1)^{-b}{}_2F_1\bigg(b, 1-a; 1+b-a; \frac{1}{1-\chi}\bigg).
\nn\\
\eea
Through Pfaff's identity, we get
\bea
(\chi-1)^{-a}{}_2F_1\bigg(a, 1-b; 1+a-b, \frac{1}{1-\chi}\bigg)&=&\chi^{-a}{}_2F_1\bigg(a, a; 1+a-b; \frac{1}{\chi}\bigg);
\nn\\
(\chi-1)^{-b}{}_2F_1\bigg(b, 1-a; 1+b-a, \frac{1}{1-\chi}\bigg)&=&\chi^{-b}{}_2F_1\bigg(b, b; 1+b-a; \frac{1}{\chi}\bigg).
\eea
Hence, we obtain
\bea
&&
(1-\chi)^{1-a-b}{}_2F_1(1-b, a; 1+a-b; 2-a-b; 1-\chi)
\nn\\
&=&
(-1)^{1-a-b}\frac{\Gamma(2-a-b)\Gamma(b-a)}{\big(\Gamma(1-a)\big)^2}
\chi^{-a}{}_2F_1\bigg(a, a; 1+a-b; \frac{1}{\chi}\bigg)
\nn\\
&&+
(-1)^{1-a-b}
\frac{\Gamma(2-a-b)\Gamma(a-b)}{\big(\Gamma(1-b)\big)^2}
\chi^{-b}{}_2F_1\bigg(b, b; 1+b-a; \frac{1}{\chi}\bigg).
\eea
Two identities both use the OPE block \cite{Czech:2015qta,Czech:2016xec,deBoer:2016pqk} branch and its partner as a basis
\bea
{\cal F}_{\Delta_+}(\chi)=\chi^{-a}{}_2F_1\bigg(a, a; 1+a-b; \frac{1}{\chi}\bigg); \
{\cal F}_{\Delta_-}(\chi)=\chi^{-b}{}_2F_1\bigg(b, b; 1+b-a; \frac{1}{\chi}\bigg).
\eea
These functions are chosen with the correct asymptotic behavior in the coincident limit $P_1 \to P_2$ ($\chi \to \infty$)
\bea
\chi^{-\Delta_\pm/2} \sim (x_1-x_2)^{\xD_\pm}.
\eea
which is precisely the behavior of an OPE block, as it follows from the OPE expansion of two operators at $x_1$ and $x_2$ on the boundary.
In this sense, these hypergeometric functions are the correlators between an OPE block and a bulk operator at $X$.
It is also not difficult to see that they follow from the analytic continuation of the results in AdS \cite{Kabat:2016zzr}, presented as a three-point function.
\\

\noindent
Finally, we reach the following identity
\bea
I_{\gamma}(\rho)=C_+{\cal F}_{\Delta_+}(\chi)+C_-{\cal F}_{\Delta_-}(\chi),
\eea
where
\bea
C_+&=&C_1\frac{\Gamma(a+b)\Gamma(b-a)}{\big(\Gamma(b)\big)^2}
+C_2(-1)^{1-a-b}\frac{\Gamma(2-a-b)\Gamma(b-a)}{\big(\Gamma(1-a)\big)^2};
\nn\\
C_-&=&C_1\frac{\Gamma(a+b)\Gamma(a-b)}{\big(\Gamma(a)\big)^2}
+C_2(-1)^{1-a-b}\frac{\Gamma(2-a-b)\Gamma(a-b)}{\big(\Gamma(1-b)\big)^2}.
\eea
\\

\noindent
We want to take the limit
\bea
q\rightarrow\infty
\eea
to determine the coefficients $C_{1, 2}$.
This limit only happends in $\alpha\beta<0$, and the $I_{\gamma}$ is
\bea
I_{\gamma}(\rho)
=\frac{\Gamma(\Delta_+)\Gamma(\Delta_-)}{(4\pi)^{\frac{d+1}{2}}\Gamma\big(\frac{d+1}{2}\big)}
\int_{-\infty}^{\infty}d\lambda\ {}_2F_1
\bigg(\Delta_+, \Delta_-; \frac{d+1}{2}; \frac{1-\eta q\sinh\lambda}{2}\bigg).
\eea
The connection formula with $q\rightarrow\infty$ is
\bea
&&
{}_2F_1
\bigg(\Delta_+, \Delta_-; \frac{d+1}{2}; \frac{1\mp q\sinh\lambda}{2}\bigg)
\nn\\
&\sim&
\frac{\Gamma\big(\frac{d+1}{2}\big)\Gamma(\Delta_--\Delta_+)}
{\Gamma(\Delta_-)\Gamma\big(\frac{1}{2}-\frac{\Delta_+-\Delta_-}{2}\big)}
\bigg(\frac{\eta q\sinh\lambda}{2}\bigg)^{-\Delta_+}
\nn\\
&&+
\frac{\Gamma\big(\frac{d+1}{2}\big)\Gamma(\Delta_+-\Delta_-)}
{\Gamma(\Delta_+)\Gamma\big(\frac{1}{2}-\frac{\Delta_--\Delta_+}{2}\big)}
\bigg(\frac{\eta q\sinh\lambda}{2}\bigg)^{-\Delta_-}
\eea
if
\bea
\Delta_+-\Delta_-\notin \mathbb{Z}.
\eea
The relevant integral is
\bea
\int_{-\infty}^{\infty}d\lambda\ (\sinh\lambda)^{-s}
=(-i)^{-s}\int_{-\infty}^{\infty}dx\ (\cosh x)^{-s}
=(-i)^{-s}\sqrt{\pi}\frac{\Gamma\big(\frac{s}{2}\big)}{\Gamma\big(\frac{s+1}{2}\big)},
\eea
in which we define a new variable
\bea
x=\lambda+i\frac{\pi}{2},
\eea
which gives
\bea
\sinh(\lambda)=-i\cosh(x).
\eea
Therefore, we obtain
\bea
&&
I_{\gamma}(\rho)
\nn\\
&\sim&
\frac{\Gamma\big(\frac{d+1}{2}\big)\Gamma(\Delta_--\Delta_+)}
{\Gamma(\Delta_-)\Gamma\big(\frac{1}{2}-\frac{\Delta_+-\Delta_-}{2}\big)}
(2i\eta)^{\Delta_+}\sqrt{\pi}
\frac{\Gamma\big(\frac{\Delta_+}{2}\big)}
{\Gamma\big(\frac{\Delta_++1}{2}\big)}q^{-\Delta_+}
\nn\\
&&+
\frac{\Gamma\big(\frac{d+1}{2}\big)\Gamma(\Delta_+-\Delta_-)}
{\Gamma(\Delta_+)\Gamma\big(\frac{1}{2}-\frac{\Delta_--\Delta_+}{2}\big)}
(2i\eta)^{\Delta_-}\sqrt{\pi}
\frac{\Gamma\big(\frac{\Delta_-}{2}\big)}
{\Gamma\big(\frac{\Delta_-+1}{2}\big)}q^{-\Delta_-}
\nn\\
&=&
C_+q^{-\Delta_+}+C_-q^{-\Delta_-},
\eea
where
\bea
C_+&=&(2i\eta)^{\Delta_+}\sqrt{\pi}
\frac{\Gamma\big(\frac{d+1}{2}\big)\Gamma(\Delta_--\Delta_+)\Gamma\big(\frac{\Delta_+}{2}\big)}
{\Gamma(\Delta_-)\Gamma\big(\frac{1}{2}-\frac{\Delta_+-\Delta_-}{2}\big)\Gamma\big(\frac{\Delta_++1}{2}\big)};
\nn\\
C_-&=&(2i\eta)^{\Delta_-}\sqrt{\pi}
\frac{\Gamma\big(\frac{d+1}{2}\big)\Gamma(\Delta_+-\Delta_-)\Gamma\big(\frac{\Delta_-}{2}\big)}
{\Gamma(\Delta_+)\Gamma\big(\frac{1}{2}-\frac{\Delta_--\Delta_+}{2}\big)\Gamma\big(\frac{\Delta_-+1}{2}\big)}.
\eea
\\

\noindent
We can solve the $C_1$ and $C_2$ as in the following:
\bea
C_+=C_1U_++C_2V_+; \
C_-=C_1U_+C_2V_-,
\eea
where
\bea
U_+=\frac{\Gamma(a+b)\Gamma(b-a)}{\big(\Gamma(b)\big)^2}; \
U_-=\frac{\Gamma(a+b)\Gamma(a-b)}{\big(\Gamma(a)\big)^2},
\eea
and
\bea
V_+=(-1)^{1-a-b}\frac{\Gamma(2-a-b)\Gamma(b-a)}{\big(\Gamma(1-a)\big)^2}; \
V_-=(-1)^{1-a-b}\frac{\Gamma(2-a-b)\Gamma(a-b)}{\big(\Gamma(1-b)\big)^2}.
\eea
Using the duplication identity
\bea
\Gamma(2z)=\frac{2^{2z-1}}{\sqrt{\pi}}\Gamma(z)\Gamma\bigg(z+\frac{1}{2}\bigg),
\eea
the coefficients become
\bea
C_+=(i\eta)^{2a}N_1U_+; \ C_-=(i\eta)^{2b}N_1U_-,
\eea
where
\bea
N_1=\sqrt{\pi}\frac{\Gamma\big(a+b+\frac{1}{2}\big)\Gamma(a)\Gamma(b)}
{\Gamma(a+b)\Gamma\big(a+\frac{1}{2}\big)\Gamma\big(b+\frac{1}{2}\big)}.
\eea
We solve the two linear equations to show
\bea
C_1&=&\frac{C_+V_--C_-V_+}{U_+V_--U_-V_+}
=N_1\frac{(i\eta)^{2a}\frac{V_-}{U_-}-(i\eta)^{2b}\frac{V_+}{U_+}}
{\frac{V_-}{U_-}-\frac{V_+}{U_+}};
\nn\\
C_2&=&\frac{U_+C_--U_-C_+}{U_+V_--U_-V_+}
=N_1\frac{(i\eta)^{2b}-(i\eta)^{2a}}{\frac{V_-}{U_-}-\frac{V_+}{U_+}},
\eea
where
\bea
\frac{V_+}{U_+}&=&(-1)^{1-a-b}\frac{\Gamma(2-a-b)}{\Gamma(a+b)}
\frac{\big(\Gamma(b)\big)^2}{\big(\Gamma(1-a)\big)^2};
\nn\\
\frac{V_-}{U_-}&=&(-1)^{1-a-b}\frac{\Gamma(2-a-b)}{\Gamma(a+b)}
\frac{\big(\Gamma(a)\big)^2}{\big(\Gamma(1-b)\big)^2}.
\eea
If $a$ and $b$ are both not integers, we can use the following identities
\bea
\bigg(\frac{\Gamma(a)}{\Gamma(1-b)}\bigg)^2
=\bigg(\frac{\Gamma(a)\Gamma(b)\sin(\pi b)}{\pi}\bigg)^2; \
\bigg(\frac{\Gamma(b)}{\Gamma(1-a)}\bigg)^2
=\bigg(\frac{\Gamma(a)\Gamma(b)\sin(\pi a)}{\pi}\bigg)^2
\eea
to simply the expressions of $C_{1, 2}$,
\bea
C_1&=&N_1\frac{(i\eta)^{2a}\sin^2(\pi b)-(i\eta)^{2b}\sin^2(\pi a)}
{\sin^2(\pi b)-\sin^2(\pi a)};
\nn\\
C_2&=&N_1\frac{\pi^2\Gamma(a+b)\big((i\eta)^{2b}-(i\eta)^{2a}\big)}
{(-1)^{1-a-b}\Gamma(2-a-b)\big(\Gamma(a)\Gamma(b)\big)^2\sin\big(\pi(a+b)\big)\sin\big(\pi(b-a)\big)}.
\eea
Hence, the geodesic-integrated Wightman function is given by the correlators between a bulk operator and a linear combination of an OPE block and its partner.
We can still use the AdS analytical continuation to obtain the dS result.
However, we need to make a linear combination as in the Wightman function.

\section{Conformal Defects and Anomaly}
\label{sec:7}
\noindent
We consider a $d$-dimensional Euclidean CFT with a global internal symmetry group $G$ and insert a flat $p$-dimensional conformal defect $W$ (bare defect operator) along a flat submanifold $\mathbb{R}^p$ with coordinates $x=(\tau^a, x_{\perp}^j)$,
where the index for the CFT coordinates along the defect is $a=1, 2, \cdots, p$, and the index for the broken directions is $j=1, 2, \cdots, d-p$.
The defect operators are inserted at $x_{\perp}=0$.
The bare defect operator $W$ is analogous to inserting $W(1)=\exp\big(-S_{\mathrm{defect}}(\phi)\big)$ on $\mathbb{R}^p$.
One can deform the bare defect insertion $W(1)$ by turning on sources of the operator $\hat{t}(\tau)$, $w^j(\tau)$,
\bea
&&
W\bigg(e^{\int d^p\tau\ w_j(\tau)\hat{t}_j(\tau)}\bigg)
\nn\\
&\equiv&\sum_{n=0}^{\infty}\frac{1}{n!}\int d\tau_1d\tau_2\cdots d\tau_n\ w_{i_1}(\tau_1)w_{i_2}(\tau_2)\cdots w_{i_n}(\tau_n)W(1)\hat{t}_{i_1}\hat{t}_{i_2}\cdots\hat{t}_{i_n}.
\eea
\\

\noindent
The local defect \cite{Billo:2016cpy} breaks the symmetry group from $G$ to $H$.
The current conservation equation must be modified only at $x_{\perp}=0$,
\bea
\partial_{\mu}J^{\mu}_a(x)=\delta^{(d-p)}(x_{\perp})P^j_a\hat{t}_j(\tau),
\eea
where $P^j_a$ is a projector extracting the part of the generator $a$ lying in the boken direction $j$.
The tilt operator $\hat{t}_j(\tau)$ is the response that represents the infinitesimal action of the broken internal symmetry.
\\

\noindent
The generating functional for the $PT$-invariant state is
\bea
Z[r, w]=\int{\cal D}\phi\ e^{-S}W\bigg(e^{\int d^p\tau\ w_j(\tau)\hat{t}_j(\tau)}\bigg),
\eea
where $S$ is the CFT bulk action.
We can use functional derivatives with respect to $w_j$ to obtain the tilt operator.
A group element $g\in G$ acts on the sources by a map $w\mapsto L_g w$.
An infinitesimal transformation of $g=\exp(\lambda)$ with $\lambda$ in the Lie algebra is $L_{e^{\lambda}}w=w+l(\lambda, w)+{\cal O}(\lambda^2)$, where $l(\lambda, w)$ is the infinitesimal variations of the local defect couplings.
We express the local defect variation $l(\lambda, w)$ in terms of a formal power series in $w$,
\bea
l(\lambda, w)=\sum_{k=0}^{\infty}\frac{1}{k!}l_k(\lambda; w, w, \cdots, w),
\eea
where $l_k(\lambda; w, w, \cdots, w)$ is a multi-linear functional of $k$ copies of $w$.
\\

\noindent
We introduce the shorthand notation:
\bea
\hat{t}(w)\equiv\int d^p\tau\ w_j(\tau)\hat{t}_j(\tau).
\eea
The logarithm of the generating functional about the connected correlators is
\bea
\ln Z[r, w]=\sum_{n\ge 0}\frac{1}{n!}\big\langle \big(\hat{t}(w)\big)^n\big\rangle_c.
\eea
The logarithm of the partition function under the infinitesimal transformation becomes \cite{Belton:2025ief}
\bea
\ln Z[L_g w]=\ln Z[w]+ A[\lambda, w],
\eea
where $A[\lambda, w]$ is a defect-local anomaly functional, which is defined as a power series
\bea
A[\lambda, w]=\sum_{n=1}^{\infty}\frac{1}{n!}A_n[\lambda; w, w, \cdots, w].
\eea
Each $A_n[\lambda; w_1, w_2, \cdots, w_n]$ is multilinear and symmetric in the $w_j$'s.
The $A_0$ is simply the anomaly evaluated at a zero-defect source.
Since the theory with $w=0$ has an exact global symmetry $G$, that anomaly must vanish.
The $A_1$ is linear in $w$, which a local counterterm can eliminate.
The simplest case containing non-trivial anomaly data is $A_{2}$ \cite{Drukker:2025dfm}.
\\

\noindent
We expand the logarithm of the partition function up to the first order in $\lambda$,
\bea
\delta_{\lambda}\ln Z=A[\lambda, w],
\eea
where
\bea
\delta_{\lambda}\ln Z=\ln \big(Z[w+l(\lambda, w)]\big)-\ln \big(Z[w]\big).
\eea
We now vary the local defect source up to the first order in $\lambda$ to show \cite{Belton:2025ief}
\bea
\ln Z[w+l(\lambda, w)]=\sum_{n\ge 0}\frac{1}{n!}\bigg\langle\bigg(\hat{t}(w)+\hat{t}\big(l(\lambda, w)\big)\bigg)^n\bigg\rangle_c.
\eea
The local defect variation of the logarithm of the partition function is \cite{Belton:2025ief}
\bea
\delta_{\lambda}^{\mathrm{defect}}\ln Z
=\sum_{n\ge 0}\frac{1}{n!}\sum_{k\ge 0}\frac{1}{k!}\big\langle\hat{t}\big(l_k(\lambda; w, w, \cdots, w)\big)\big(\hat{t}(w)\big)^{n}\big\rangle_c.
\eea
We extract the terms with a fixed-order in $n$ for $w$ \cite{Belton:2025ief},
\bea
\frac{1}{n!}\sum_{k=0}^n\frac{n!}{k!(n-k)!}\big\langle\hat{t}\big(l_k(\lambda; w, w, \cdots, w)\big)\big(\hat{t}(w)\big)^{n-k}\big\rangle_c.
\eea
Hence, we obtain the integral identities for the connected correlators \cite{Belton:2025ief,Drukker:2025dfm}
\bea
\sum_{k=0}^n\frac{n!}{k!(n-k)!}\big\langle\hat{t}\big(l_k(\lambda; w, w, \cdots, w)\big)\big(\hat{t}(w)\big)^{n-k}\big\rangle_c
=A_n[\lambda; w, w, \cdots, w].
\eea
\\

\noindent
For $n=0$, it provides \cite{Belton:2025ief}
\bea
\int d^p\tau\ \langle \hat{t}_j(\tau)\rangle_c=0.
\label{id1}
\eea
When $n=1$, we get an integral constraint on the two-point tilt function \cite{Belton:2025ief}
\bea
\int d^p\tau_1\ \langle\hat{t}_j(\tau_1)\hat{t}_k(\tau_2)\rangle_c=0.
\label{id2}
\eea
For $n=2$, the integrated insertion of a broken charge (the first term) is equal to the transformation of the two-point function under the nonlinear change of coordinates induced by $l_1$ \cite{Belton:2025ief,Drukker:2025dfm}.
\bea
&&
\bigg(\int d^p\tau_1\ \langle t_j(\tau_1)t_k(\tau_2)t_l(\tau_3)\rangle_c\bigg)
\nn\\
&&
+l_1{}^m(e_j; e_k)\langle t_m(\tau_2)t_l(\tau_3)\rangle_c
+l_1{}^m(e_j; e_l)\langle t_k(\tau_2)t_m(\tau_3)\rangle_c
\nn\\
&=&\frac{1}{2}\frac{\delta^2}{\delta w_k(\tau_2)\delta w_l(\tau_3)}A_2[e_j. w, w]\bigg|_{w=0}.
\eea
For $n=2$, the anomaly term has a possible contribution from the coincident points $\tau_2=\tau_3$, but we drop it for separated points $\tau_2\neq\tau_3$ \cite{Drukker:2025dfm}.
\\

\noindent
Conformal symmetry makes it easier to determine correlators.
Therefore, we consider the infinite planar defect with the full conformal group for CFT$_d$ is SO($d+1$, $1$).
Introducing a flat $p$-dimensional local defect preserves a conformal group along the defect, SO($p+1$, 1).
We have the symmetry for translations along the defect, rotations within the defect (SO($p$)), rotations in directions normal to the defect (SO($d-p$)), dilatations, and the special conformal transformations tangent to the defect.
Therefore, the preserved symmetry group is SO($p+1, 1$)$\times$SO($d-p$), which is called the "defect conformal group" \cite{Billo:2016cpy}.
We consider the tilt operator $\hat{t}_j$ that breaks an internal symmetry $G\rightarrow H$.
The tilt operator has a scaling dimension $\Delta_{\hat{t}}=p>0$.
It transforms as a vector under the broken symmetry.
\\

\noindent
The translational symmetry constrains the expectation value of the one-point tilt operator to be a constant.
We then use the scaling symmetry to restrict the constant to zero.
Therefore, the one-point tilt operator satisfies the integral constraint for $n=0$ \eqref{id1} \cite{Belton:2025ief}.
\\

\noindent
We later determine the two-point tilt operator, including the coincident points, and also the three-point tilt operator.
We then use the tilt correlators to discuss the anomaly term $A_2$.
Finally, we provide the bulk interpretation to the tilt correlators.

\subsection{Two-Point}
\noindent
For the two-point tilt operators, we can first consider translational and rotational symmetries along the defect directions to show that the two-point Green's function of the tile operators depends only on the distance in defect coordinates.
We then apply the scaling to the two-point tilt operators for the separated points, which implies
\bea
\langle\hat{t}_j(\tau_1)\hat{t}_k(\tau_2)\rangle_c=\frac{C_{jk}}{|\tau_{12}|^{2p}}, \ \tau_1\neq\tau_2
\eea
where
\bea
\tau_{12}\equiv\tau_1-\tau_2.
\eea
Because the tile operators transform as a vector under the internal representation, and the vacuum state is invariant under the internal symmetry, the only possible invariant tensor is proportional to the identity matrix
\bea
C_{jk}=C_t\delta_{jk}.
\eea
\\

\noindent
For the coincident points, the two-point tilt operator has an additional term from a finite sum of derivatives of the delta function, and it scales as
\bea
\partial_{\tau}^n(\alpha\tau)=\alpha^{-n-1}\partial_{\tau}^n\delta(\tau).
\eea
Because the tilt operator has the conformal dimension $p$, the scaling dimension of the derivative of the delta function needs to be $2p$,
\bea
n+1=2p.
\eea
In other words, the two-point tilt operators, including the coincident points, are
\bea
\langle\hat{t}_j(\tau_1)\hat{t}_k(\tau_2)\rangle_c=
\frac{C_t \delta_{jk}}{|\tau_{12}|^{2p}}
+A_{jk}\partial_{\tau_1}^{2p-1}\delta(\tau_{12}).
\eea
Because the tilt operators are the bosonic operators, we can impose exchange symmetry
\bea
\langle\hat{t}_j(\tau_1)\hat{t}_k(\tau_2)\rangle_c=\langle\hat{t}_k(\tau_2)\hat{t}_j(\tau_1)\rangle_c.
\eea
The $\partial_{\tau_1}^{2p-1}\delta(\tau_{12})$ is odd under $\tau_{12} \rightarrow -\tau_{12}$.
Exchanging symmetry gives
\bea
A_{jk}=-A_{kj}.
\eea
The $A_{jk}$ is antisymmetric.
Because the tilt operator lives in the transverse vector representation, the defect preserves the transverse rotating group SO($d-p$).
The only invariant two-index tensor is $\delta_{jk}$.
Therefore, there is no invariant antisymmetric tensor $A_{jk}$.
Therefore, we get $A_{jk}=0$.
Hence, the two-point tilt function, including the coincident point case, is \cite{Belton:2025ief}
\bea
\langle\hat{t}_j(\tau_1)\hat{t}_k(\tau_2)\rangle_c=\frac{C_{jk}}{|\tau_{12}|^{2p}}.
\eea
This result respects the $PT$-invariance \cite{Huang:2025gmq}.
\\

\noindent
The two-point tilt function has a divergence near $\tau_{12}=0$.
We provide a test function $f(\tau_{12})$ for defining the first term.
The expansion of $f(\tau_{12})$ near $\tau_{12}=0$ gives
\bea
f(\tau)=f(0)+\tau f^{\prime}(0)+\cdots,
\eea
where $\tau\equiv\tau_{12}$.
Because $1/|\tau|^{2p}$ is even, only the even powers contribute to the divergent part
\bea
\sum_{n=0}^{p-1}\frac{f^{(2n)}(0)}{(2n)!}\int_{\epsilon<|\tau|}d\tau\ \frac{\tau^{2n}}{|\tau|^{2p}}
=
2\sum_{n=0}^{p-1}\frac{f^{(2n)}(0)}{(2n)!}\frac{\epsilon^{-(2p-2n-1)}}{2p-2n-1},
\eea
in which we use the integration result
\bea
\int_{\epsilon<|\tau|}d\tau\ \frac{\tau^{2n}}{|\tau|^{2p}}=\frac{2}{2p-2n-1}\epsilon^{-(2p-2n-1)}+\mathrm{finite}, \ n=0, 1, \cdots, p-1.
\eea
Hence, the finite part is given by
\bea
\int d\tau\ \langle\hat{t}_j(\tau)\hat{t}_k(0)\rangle_cf(\tau)
\sim C_t\delta_{jk}\lim_{\epsilon\rightarrow 0}
\bigg(\int_{|\tau|>\epsilon}d\tau\ \frac{f(\tau)}{|\tau|^{2p}}
-2\sum_{n=0}^{p-1}\frac{f^{(2n)}(0)}{(2n)!}\frac{\epsilon^{-(2p-2n-1)}}{2p-2n-1}\bigg).
\nn\\
\eea
When substituting
\bea
f(\tau)=1,
\eea
it shows the $n=1$ integral identity \eqref{id2}.
\\

\noindent
We can use the same method to analyze the logarithmic divergence term.
Using the $m$-th Taylor term of $f(\tau)$ gives the following relevant integration
\bea
\int dr\ r^{p-1}\frac{r^m}{r^{2p}}=\int dr\ r^{m-p-1},
\eea
in which we use
\bea
d^p\tau=r^{p-1}drd\Omega_{p-1}.
\eea
The logarithmic divergence occurs only when
\bea
m-p-1=-1.
\eea
Because the angular integral vanishes when $m$ is odd, the logarithmic divergence appears only if $p$ is even.
The logarithmic divergence implies scale dependence or a Weyl anomaly.
The Weyl anomaly contributes to $A_2$ for the even $p$.

\subsection{Three-Point}
\noindent
For $p>1$, the $\tau_{12}$ is a vector.
The unique scalar conformal structure is given by $|\tau_{12}|$.
Therefore, the coordinate dependence in three-point tilt function $\langle t_j(\tau_1)t_k(\tau_2)t_l(\tau_3)\rangle_c$ is symmetric under permutuations of $\tau_{1, 2, 3}$.
Any antisymmetric invariant tensor would force antisymmetry under index exchange.
However, the tilt operators are bosonic, so their separated-point correlator must be symmetric under this exchange.
Therefore, the three-point tilt function vanishes for $p>1$ \cite{Drukker:2025dfm} because the antisymmetric index structures cannot be matched with symmetric coordinate dependence.
In other words, the anomaly term $A_2$ vanishes for $p>1$.
\\

\noindent
For $p=1$, however, it allows another conformal structure with the antisymmetric tensor, $f_{jkl}$:
\bea
\langle t_j(\tau_1)t_k(\tau_2)t_l(\tau_3)\rangle_c=\langle t_j(\tau_1)t_k(\tau_2)t_l(\tau_3)\rangle=\frac{if_{jkl}}{\tau_{12}\tau_{23}\tau_{31}}.
\eea
Under exchange of two points, the spacetime factor is antisymmetric
\bea
\tau_{12}\tau_{23}\tau_{31} \longleftrightarrow -\tau_{12}\tau_{23}\tau_{31}.
\eea
The $f_{jki}$ is also antisymmetric.
Therefore, the bosonic exchange symmetry is only respected for $p=1$ in the non-trivial three-point tilt function.
Because we consider the $PT$-invariant state, the three-point tilt function needs to respect the $PT$-invariance, which reverses the
$\tau_{12}$, $\tau_{23}$, and $\tau_{31}$ with the complex conjugate.
Therefore, we need the $i$ in the three-point tilt function to have $PT$-symmetry.
\\

\noindent
When performing the integration of $\tau_1$, the integrand of the three-point tilt function has poles at
\begin{itemize}
\item{$\tau_1=\tau_2$;}
\item{$\tau_1=\tau_3$.}
\end{itemize}
As an ordinary function, the integral is ill-defined.
In QFT, such objects are defined by a prescription (principal value/i0 prescription), i.e., as a distribution acting on test functions.
Therefore, we study it as a distribution $I(\tau_2, \tau_3)$ defined by
\bea
I(\tau_1, \tau_2, \tau_3)\equiv\int d\tau_1\ \frac{1}{(\tau_{12})(\tau_{23})(\tau_{31})}.
\eea
Factor out the $\tau_{23}$ because it does not depend on $\tau_1$,
\bea
I(\tau_2, \tau_3)=\frac{1}{\tau_{23}}\int d\tau_1\ \frac{1}{(\tau_{12})(\tau_{31})}
=-\frac{1}{\tau_{23}^2}\int d\tau_1\ \bigg(\frac{1}{\tau_{12}}+\frac{1}{\tau_{31}}\bigg).
\eea
\\

\noindent
If we define each $1/(\tau_1-a)$ by principal value on $\mathbb{R}$,
\bea
\mathrm{PV}\int^{\infty}_{-\infty}\frac{d\tau_1}{\tau_1-a}=0
\eea
because it is an odd integrand after shifting
\bea
\tau_1\rightarrow\tau_1+a.
\eea
For $\tau_2\neq\tau_3$, the two PV integrals vanish separately, giving
\bea
I(\tau_2, \tau_3)=0.
\eea
Even though the PV integral vanishes for $\tau_2\neq\tau_3$, it becomes singular when $\tau_2\rightarrow\tau_3$ because the two poles in $\tau_1$ collide and produce a distribution.
To compute that distribution cleanly, use the standard QFT identity (this is the basic distributional definition of $1/x$ with an i0 prescription):
\bea
\frac{1}{x\pm i0}=\mathrm{PV}\bigg(\frac{1}{x}\bigg)\mp i\pi\delta(x).
\label{i0p}
\eea
We now regulate the integrand by assigning an i0 prescription to each factor.
This important structural point is that the product of two such factors produces a delta -function when poles collide.
\\

\noindent
Let us focus on the core piece
\bea
J(\tau_2, \tau_3)&\equiv&\int d\tau_1\ \frac{1}{(\tau_1-\tau_2\pm i0)(\tau_1-\tau_3\mp i0)}
\nn\\
&=&\int d\tau_1\ \frac{1}{\tau_2-\tau_3\mp 2i0}
\bigg(\frac{1}{\tau_1-\tau_2\pm i0}
-\frac{1}{\tau_1-\tau_3\mp i0}\bigg).
\eea
The regularization independent piece is
\bea
J(\tau_2, \tau_3)&\sim&\int d\tau_1\ \frac{1}{\tau_2-\tau_3\mp 2i0}
\big(\mp i\pi\delta(\tau_1-\tau_2)\mp i\pi\delta(\tau_1-\tau_3)\big)
\nn\\
&=&\frac{\mp 2\pi i}{\tau_2-\tau_3\mp 2i0}
\nn\\
&=&2\pi^2\delta(\tau_2-\tau_3).
\eea
The regularization independent piece of $J$ produces a delta function $\delta(\tau_{23})$.
\\

\noindent
Recall that
\bea
I(\tau_2, \tau_3)=\frac{1}{\tau_{23}}\int d\tau_1\ \frac{1}{\tau_{12}\tau_{31}}=\frac{1}{\tau_{23}}J(\tau_2, \tau_3),
\eea
but $J$ contains $\delta(\tau_{23})$.
Therefore, we get
\bea
I(\tau_2, \tau_3)\sim 2\pi^2 \frac{\delta(\tau_{23})}{\tau_{23}}=-2\pi^2\delta^{\prime}(\tau_{23}),
\eea
in which we use the following identity
\bea
\frac{\delta(x)}{x}\equiv-\delta^{\prime}(x),
\eea
which is a consequence of $x\delta(x)=0$ with differentiateing both sides to get
\bea
\delta(x)+x\delta^{\prime}(x)=0\Rightarrow x\delta^{\prime}(x)=-\delta(x).
\eea
\\

\noindent
We multiply the sources by the integrated 3-point function
\bea
i\int d\tau_2d\tau_3\ f_{jkl}\lambda_jw_k(\tau_2)w_l(\tau_3)I(\tau_2, \tau_3)
&\sim&-2i\pi^2\int d\tau_2d\tau_3\ f_{jkl}\lambda_jw_k(\tau_2)w_l(\tau_3)\delta^{\prime}(\tau_{23})
\nn\\
&=&2i\pi^2\int d\tau\ f_{jkl}\lambda_jw_k(\tau)\partial_{\tau}w_l(\tau),
\eea
in which we use the following identity
\bea
\int d\tau_3\ F(\tau_3)\delta^{\prime}(\tau_2-\tau_3)=-\frac{d}{d\tau_2}F(\tau_2).
\eea
This is exactly the local "Wess-Zumino-type" structure \cite{Drukker:2025dfm}
\bea
A_2[\lambda; w, w]\sim 2i\pi^2 \int d\tau\ f_{jkl}\lambda_jw_k\partial_{\tau}w_l.
\eea
The $w^j(\tau)$ are the local coordinates on $G/H$ describing fluctuations away from the local defect in broken-symmetry directions, where
\begin{itemize}
\item{The symmetry group is $G$;}
\item{The unbroken subgroup is $H$;}
\item{The space of broken directions is the coset space $G/H$.}
\end{itemize}
The anomaly is consistent if and only if the 3-form
\bea
w=\frac{1}{6}f_{jkl}dw^j\wedge dw^k\wedge dw^l
\eea
is closed but not exact \cite{Drukker:2025dfm}.
This statement is exactly $[f_{jkl}]\in H^3(G/H, \mathbb{R})$ \cite{Drukker:2025dfm}, where
\bea
H^3(G/H, \mathbb{R})=\frac{\mathrm{closed\ 3\ forms}}{\mathrm{exact\ 3\ forms}}.
\eea
\\

\noindent
A 3-form $w$ is exact if
\bea
w=d\beta
\eea
for some 2-form $\beta$.
Physically:
\begin{itemize}
\item{Adding a local counterm built from $\beta$ changes the anomaly by $d\beta$.}
\item{Exact terms can be removed by redefining the theory.}
\end{itemize}
If $w$ is closed but not exact:
\begin{itemize}
\item{It cannot be written as $d$(something local);}
\item{Therefore, it cannot be removed.}
\end{itemize}
Therefore,
\begin{itemize}
\item{Closed$\Rightarrow$ consistent anomaly;}
\item{Exact$\Rightarrow$ removable;}
\item{Closed but not exat$\Rightarrow$ genunie anomaly.}
\end{itemize}
Hence, the line-defect tilt anomalies are classified by $H^3(G/H, \mathbb{R})$ \cite{Drukker:2025dfm}.
Coincident-point singularities generate local terms; cohomology tells us which of those local terms are genuine anomalies (i.e., cannot be removed by counterterms).
Suppose coincident points generate a new term that can be canceled by adding a local counterterm.
Then this is not an anomaly-it is scheme-dependent.
Coincident points generate candidate anomaly terms, but most of them are trivial.

\subsection{Bulk Interpretation}
\noindent
In the bulk gravity, the line defect becomes a worldline object $\gamma$.
Its internal orientation is promoted to a worldline field
\bea
u(\lambda)=\exp\big(\pi^j(\lambda)e_j)\big)\equiv \exp(\Pi)\in G/H,
\eea
where the $\pi^j(\lambda)$ are the Goldstone coordinates on the coset.
The Maurer-Cartan one-form projected to the broken directions
\bea
u^{-1}du=W^je_j+w^{\alpha}h_{\alpha},
\eea
where the broken-symmetry index is $j=1, 2, \cdots, \dim(G/H)$, and the unbroken-symmetry index is $\alpha=1, 2, \cdots, \dim(H)$.
\\

\noindent
We define
\bea
F(s) =e^{-s\Pi}de^{s\Pi}.
\eea
Therefore, we get
\bea
F(1)=\int_0^1 ds\ \frac{dF(s)}{ds}=\int_0^1ds\ e^{-s\Pi}(d\Pi)e^{s\Pi}.
\eea
We then obtain the expansion result
\bea
F(1)=d\Pi-\frac{1}{2}\lbrack\Pi, d\Pi\rbrack+\cdots.
\eea
Therefore, the expansion result gives
\bea
W^j=d\pi^k-\frac{1}{2}f_{kl}{}^j\pi^kd\pi^l+\cdots,
\eea
where
\bea
\lbrack e_j, e_k\rbrack=f_{jk}{}^le_l+f_{jk}{}^{\alpha}h_{\alpha}.
\eea
\\

\noindent
The worldline effective action corresponding to the two-point tilt function is
\bea
S_2=\frac{\kappa}{2}\int_{\gamma}d\lambda\ C_t\delta_{jk}W^jW^k
\approx\frac{\kappa}{2}\int_{\gamma}d\lambda\ g_{jk}
(\partial_{\lambda}\pi^j)(\partial_{\lambda}\pi^k)+{\cal O}(\pi^4),
\eea
where $\gamma$ is a timelike (spacelike) geodesic for the dS (AdS) background.
This quadratic term is a sigma model on $G/H$.
The Goldstone fields $\pi^j$ describe fluctuations of the defect orientation.
Hence, the two-point tilt function corresponds to the kinetic energy of defect fluctuations.
The tilt operator is analogous to transverse fluctuations of a D-brane.
The effective action corresponding to the three-point tilt function is
\bea
S_3=ik\int_{B_2}B_2(W)=i\frac{k}{6}\int_{B_2}f_{jkl}\pi^jd\pi^k\wedge d\pi^j+{\cal O}(\pi^4),
\eea
where
\bea
dB_2(W)=\frac{1}{3!}f_{jkl}W^j\wedge W^k\wedge W^l.
\eea
The two-dimensional surface $B_2$ with the boundary
\bea
\partial B_2=\gamma.
\eea
The $dB_2(W)$ plays the role of an NS-NS flux in string theory.
The $f_{jkl}$ is the analog of a background NS flux on $G/H$.

\section{Discussion and Conclusion}
\label{sec:8}
\noindent
In this work, we have investigated the Euclidean CFT dual description of Lorentzian de Sitter spacetime through timelike geodesics. Starting from the analytic continuation of global AdS to the doubled static-patch de Sitter geometry \cite{Strominger:2001pn,Doi:2024nty,Huang:2025gmq}, we considered scalar fields with generic masses and established the corresponding CFT description. Because the de Sitter isometry group is identified with the conformal symmetry group on the boundary, the bulk-coordinate representation induces a non-standard adjoint operation.
This modification naturally introduces a global $PT$-defect operator that twists the inner product.
Although this adjoint structure differs from that of conventional AdS/CFT, it is precisely the ingredient required to reproduce the Bunch--Davies Wightman function \cite{Anninos:2011ui} from the CFT.
\\

\noindent
The resulting $PT$-twisted inner product provides a Hilbert-space framework for studying quantum-information quantities in dS/CFT \cite{Bender:1998ke,Bender:2002vv,Mostafazadeh:2001jk,Mostafazadeh:2001nr,Mostafazadeh:2002id,Mostafazadeh:2002maq}. Within this framework, we computed the entanglement entropy of CFT$_2$ with complex central charge and obtained a real-valued result.
We found that the entropy is sensitive only to the real part of the central charge, corresponding to the one-loop contribution in the bulk gravitational description. 
In contrast, the classical gravitational contribution remains invisible to this probe.
\\

\noindent
To access the missing information, we introduced a single timelike geodesic.
We showed that the geodesic-integrated Wightman function is dual to the correlators between a bulk operator and a linear combination of an OPE block \cite{Czech:2015qta,Czech:2016xec,deBoer:2016pqk} and its partner. 
This construction introduces a new geometric observable in de Sitter holography. 
It suggests that timelike geodesics play a role analogous to that of spacelike geodesics in AdS/CFT.
Together with the HKLL reconstruction \cite{Xiao:2014uea}, these results provide evidence that several key structures familiar from AdS/CFT persist in the analytically continued dS/CFT framework, albeit in a modified form involving partner contributions.
A complete analog of the Radon-transform description \cite{Czech:2015qta,Czech:2016xec,deBoer:2016pqk} relating HKLL reconstruction and OPE blocks remains absent and deserves further investigation.
\\

\noindent
A crucial ingredient in our construction is the doubled static-patch geometry obtained through analytic continuation.
Unlike the conventional static patch, the doubled geometry possesses an antipodal $\mathbb{Z}_2$ symmetry, which underlies the $PT$-symmetric inner product.
The fact that this structure emerges for scalar fields with generic masses suggests that it reflects a fundamental property of the analytically continued dS/CFT correspondence rather than an accidental feature of a particular model.
It would therefore be interesting to generalize the construction to fields with spin and determine whether the corresponding matter content modifies the inner-product structure or the realization of the $PT$ symmetry.
\\

\noindent
We also investigated conformal defects \cite{Billo:2016cpy} through the two-point and three-point tilt operators.
The associated anomaly can be understood through an integral identity, and the resulting effective action admits a natural bulk interpretation.
In particular, the tilt operator behaves analogously to a transverse fluctuation mode of a brane. 
At the same time, the effective action resembles a combination of a brane worldvolume action and a coupling to background flux.
Although de Sitter space lacks the supersymmetric and BPS structures familiar from AdS, the $PT$-symmetric framework may provide an alternative mechanism for controlling instabilities and understanding non-Hermitian sectors of the theory.
Extending HKLL reconstruction and OPE-block constructions \cite{Czech:2015qta,Czech:2016xec,deBoer:2016pqk,Huang:2019wzc,Huang:2020cye} to include local defect operators may therefore provide a useful route toward identifying stable extended objects and their holographic dual descriptions in de Sitter space.
\\

\noindent
Several open questions remain.
Interacting quantum field theories in de Sitter space are not generically unstable. 
However, light fields in the complementary series can exhibit strong infrared effects that invalidate naive perturbation theory.
Determining how the $PT$-symmetric Hilbert-space construction extends to interacting theories is therefore an important future direction.
More broadly, understanding the roles of timelike geodesics, conformal defects \cite{Billo:2016cpy}, and quantum information observables \cite{Nakata:2020luh,Doi:2022iyj} in de Sitter holography may help establish a more complete dS/CFT dictionary and provide new insights into quantum gravity in cosmological spacetimes.

\section*{Acknowledgments}
\noindent 
We thank Wu-zhong Guo and Tadashi Takayanagi for their helpful discussion.
XH acknowledges the NSFC Grants (Grants No. 12247103 and No. 12475072).
CTM thanks Nan-Peng Ma for his encouragement.

%\noindent
%The author acknowledges the YST Program of the APCTP; 
%Post-Doctoral International Exchange Program (Grant No. YJ20180087); 
%China Postdoctoral Science Foundation, Postdoctoral General Funding: Second Class (Grant No. 2019M652926); 
%Foreign Young Talents Program (Grant No. QN20200230017); 
%Science and Technology Program of Guangzhou (Grant No. 2019050001).

\appendix
\section{Derivation of Some Useful Identities}
\label{sec:A}
\noindent
We present useful identities that help the timelike integrated-geodesic Wightman function satisfy the hypergeometric equation.
The first identity is
\bea
(1-x){}_2F_1\bigg(1-s, \frac{3}{2}; \frac{3}{2}-s; x\bigg)
=-(2s-1)y_s(x)
+2sy_{s-1}(x), \ |x|\le 1,
\eea
where
\bea
y_s(x)={}_2F_1\bigg(-s, \frac{1}{2}; \frac{1}{2}-s; x\bigg).
\eea
We then use the first identity to derive the second identity
\bea
qP_s^{\prime}=sP_s-\frac{s^2}{1-2s}P_{s-1}, \ 0<|q|\le 1,
\eea
where
\bea
P_s(q)=\bigg(-\frac{1+q}{2}\bigg)^s{}_2F_1\bigg(-s, \frac{1}{2}; \frac{1}{2}-s; \frac{1-q}{1+q}\bigg).
\eea
The third identity is
\bea
C(s)\frac{(2s+d-1)s^2}{1-2s}=C(s-1)(s+\Delta_+-1)(s+\Delta_--1),
\eea
where
\bea
C(s)\equiv\frac{2\sqrt{\pi}}{(4\pi)^{\frac{d+1}{2}}}\frac{\Gamma(\Delta_++s)\Gamma(\Delta_-+s)\big(\Gamma(-s)\big)^2}
{\Gamma\big(\frac{d+1}{2}+s\big)\Gamma\big(\frac{1}{2}-s\big)}.
\eea
The $\Delta_{\pm}$ are the CFT$_d$ conformal dimensions, and we have
\bea
\Delta_++\Delta_-=d.
\eea

\subsection{First Identity}
\noindent
We first let
\bea
A_1(x)={}_2F_1\bigg(1-s, \frac{3}{2}; \frac{3}{2}-s; x\bigg),
\eea
and then expand for $|x|<1$
\bea
A_(x)=\sum_{n=0}^{\infty}A_{1, n}x^n
\eea
with
\bea
A_{1, n}=\frac{(1-s)_n\big(\frac{3}{2}\big)_n}{\big(\frac{3}{2}-s\big)_nn!}.
\eea
We then have
\bea
(1-x)A_1(x)=A_{1, 0}+\sum_{n=1}^{\infty}(A_{1, n}-A_{1, n-1})x^n.
\eea
Now, we write
\bea
y_s(x)=\sum_{n=0}^{\infty}A_{2, n}x^n; \
y_{s-1}=\sum_{n=0}^{\infty}A_{3, n}x^n,
\eea
where
\bea
A_{2, n}=\frac{(-s)_n\big(\frac{1}{2}\big)_n}{\big(\frac{1}{2}-s\big)_nn!}; \
A_{3, n}=\frac{(1-s)_n\big(\frac{1}{2}\big)_n}{\big(\frac{3}{2}-s\big)_sn!}.
\eea
Hence, it is equivalent to deriving
\bea
A_{1, 0}+\sum_{n=1}^{\infty}(A_{1, n}-A_{1, n-1})x^n=\sum_{n=0}^{\infty}\big(-(2s-1)A_{2, n}+2sA_{3, n}\big)x^n.
\eea
For $n=0$, we get that
\bea
A_{1, 0}=1=-(2s-1)A_{2, 0}+2sA_{3, 0}.
\eea
For $n\ge 1$, we express everytinh in terms of $A_{3, n}$.
First, we have
\bea
A_{1, n}=(2n+1)A_{3, n}
\eea
because we have
\bea
\frac{\big(\frac{3}{2}\big)_n}{\big(\frac{1}{2}\big)_n}=2n+1.
\eea
We also have
\bea
A_{1, n-1}=A_{3, n}\frac{n(2n+1-2s)}{n-s}.
\eea
Therefore, we obtain
\bea
A_{1, n}-A_{1, n-1}=A_{3, n}\bigg((2n+1)-\frac{n(2n+1-2s)}{n-s}\bigg).
\eea
The numerator is
\bea
(2n+1)(n-s)-n(2n+1-2s)=-s,
\eea
so that we get
\bea
A_{1, n}-A_{1, n-1}=-\frac{s}{n-s}A_{3, n}.
\eea
We also have
\bea
A_{2, n}=A_{3, n}\frac{2s\big(n+\frac{1}{2}-s\big)}{(2s-1)(n-s)}.
\eea
Therefore, we get
\bea
-(2s-1)A_{2, n}+2sA_{3, n}=
-\frac{2s\big(n+\frac{1}{2}-s\big)}{n-s}A_{3, n}
=-\frac{s}{n-s}A_{3, n}.
\eea
Thus, we show
\bea
A_{1, n}-A_{1, n-1}=-(2s-1)A_{2, n}+2sA_{3, n}
\eea
for every $n\ge 1$.
Therefore, we prove
\bea
(1-x){}_2F_1\bigg(1-s, \frac{3}{2}; \frac{3}{2}-s; x\bigg)
=-(2s-1)y_s(x)
+2sy_{s-1}(x)
\eea
when $|x|<1$.
For $|x|=1$, the identity still holds by the analytical continuation.

\subsection{Second Identity}
\noindent
We define
\bea
x\equiv\frac{1-q}{1+q},
\eea
and then we get
\bea
q=\frac{1-x}{1+x}.
\eea
Therefore, we get
\bea
\frac{1+q}{2}=\frac{1}{1+x},
\eea
so that we have
\bea
P_s(q)=(-1)^s(1+x)^{-s}y_s(x),
\eea
where
\bea
y_s(x)={}_2F_1\bigg(-s, \frac{1}{2}; \frac{1}{2}-s; x\bigg).
\eea
Similarly, we also have
\bea
P_{s-1}(q)=(-1)^{s-1}(1+x)^{1-s}y_{s-1}(x)
\eea
with
\bea
y_{s-1}(x)={}_2F_1\bigg(1-s, \frac{1}{2}; \frac{3}{2}-s; x\bigg).
\eea
\\

\noindent
Now, we have
\bea
\frac{dx}{dq}=-\frac{2}{(1+q)^2}=-\frac{(1+x)^2}{2},
\eea
so that we get
\bea
q\frac{d}{dq}=-\frac{1-x^2}{2}\frac{d}{dx}.
\eea
Therefore, we get
\bea
qP_s^{\prime}=-\frac{1-x^2}{2}\frac{d}{dx}\big((-1)^s(1+x)^{-s}y_s\big).
\eea
We then compute the derivative
\bea
\frac{d}{dx}\big((1+x)^{-s}y_s\big)
=-s(1+x)^{-s-1}y_s
+(1+x)^{-s}y_s^{\prime}.
\eea
Thus, we obtain
\bea
qP_s^{\prime}=(-1)^s(1+x)^{-s}\bigg(
\frac{s}{2}(1-x)y_s-\frac{1}{2}(1-x^2)y_s^{\prime}\bigg).
\eea
We now use
\bea
y_s^{\prime}=\frac{(-s)\big(\frac{1}{2}\big)}{\frac{1}{2}-s}
{}_2F_1\bigg(1-s, \frac{3}{2}; \frac{3}{2}-s; x\bigg).
\eea
Because we have
\bea
\frac{(-s)\big(\frac{1}{2}\big)}{\frac{1}{2}-s}=-\frac{s}{1-2s},
\eea
we get
\bea
y_s^{\prime}=-\frac{s}{1-2s}{}_2F_1\bigg(1-s, \frac{3}{2}; \frac{3}{2}-s; x\bigg).
\eea
We now use the first identity,
\bea
(1-x){}_2F_1\bigg(1-s, \frac{3}{2}; \frac{3}{2}-s; x\bigg)
=-(2s-1)y_s(x)
+2sy_{s-1}(x), \ |x|\le 1,
\eea
to show
\bea
(1-x)y_s^{\prime}=-sy_s+\frac{2s^2}{2s-1}y_{s-1},
\eea
which equivalently implies
\bea
-\frac{1}{2}(1-x^2)y_s^{\prime}
=\frac{s}{2}(1+x)y_s-\frac{s^2}{2s-1}(1+x)y_{s-1}.
\eea
Hence, we now obtain
\bea
qP_s^{\prime}&=&(-1)^s(1+x)^{-s}
\bigg(\frac{s}{2}(1-x)y_s
+\frac{s}{2}(1+x)y_s
-\frac{s^2}{2s-1}(1+x)y_{s-1}\bigg)
\nn\\
&=&s(-1)^s(1+x)^{-s}y_s
-\frac{s^2}{2s-1}(-1)^s(1+x)^{1-s}y_{s-1}
\nn\\
&=&
sP_s+\frac{s^2}{2s-1}P_{s-1}.
\eea
We provide the identity
\bea
qP_s^{\prime}=sP_s-\frac{s^2}{1-2s}P_{s-1}.
\eea

\subsection{Third Identity}
\noindent
We compute the ratio
\bea
&&
\frac{C(s)}{C(s-1)}
\nn\\
&=&\frac{\Gamma(\Delta_++s)}{\Gamma(\Delta_++s-1)}
\frac{\Gamma(\Delta_-+s)}{\Gamma(\Delta_-+s-1)}
\bigg(\frac{\Gamma(-s)}{\Gamma(-s+1)}\bigg)^2
\nn\\
&&\times
\frac{\Gamma\big(\frac{d+1}{2}+s-1\big)}{\Gamma\big(\frac{d+1}{2}+s\big)}
\frac{\Gamma\big(\frac{3}{2}-s\big)}{\Gamma\big(-s+\frac{1}{2}\big)}.
\eea
We now use
\bea
\Gamma(z+1)=z\Gamma(z)
\eea
to get
\bea
&&
\frac{\Gamma(\Delta_++s)}{\Gamma(\Delta_++s-1)}=\Delta_++s-1; \
\frac{\Gamma(\Delta_-+s)}{\Gamma(\Delta_-+s-1)}=\Delta_-+s-1;
\nn\\
&&
\bigg(\frac{\Gamma(-s)}{\Gamma(-s+1)}\bigg)^2=\frac{1}{s^2}; \
\frac{\Gamma\big(\frac{d+1}{2}+s-1\big)}{\Gamma\big(\frac{d+1}{2}+s\big)}=\frac{1}{\frac{d+1}{2}+s-1};
\nn\\
&&
\frac{\Gamma\big(\frac{3}{2}-s\big)}{\Gamma\big(-s+\frac{1}{2}\big)}=-\frac{2s-1}{2}.
\eea
We then put everything together
\bea
\frac{C(s)}{C(s-1)}&=&(s+\Delta_+-1)(s+\Delta_--1)\frac{1}{s+\frac{d+1}{2}-1}\frac{1}{s^2}\bigg(-\frac{2s-1}{2}\bigg)
\nn\\
&=&(s+\Delta_+-1)(s+\Delta_--1)\bigg(-\frac{2s-1}{2\big(s+\frac{d+1}{2}-1\big)s^2}\bigg),
\eea
which equivalently implies
\bea
C(s)\frac{(2s+d-1)s^2}{1-2s}=C(s-1)(s+\Delta_+-1)(s+\Delta_--1).
\eea

\end{document}